\documentclass[10pt]{article}
\input epsf
\usepackage[dvips]{graphicx}
\usepackage{amsmath,amssymb,latexsym}
\newcommand{\Scri}{\mathcal{I}}
\newcommand{\amp}{P}
\newcommand{\width}{\mathcal{\sigma}}
\newcommand{\cosn}{n}
\newcommand{\ro}{r_0}
\newcommand{\rb}{\bar{r}}
\newcommand{\tb}{\bar{t}}
\newcommand{\constant}{\hbox{{\rm constant}}}
\newcommand{\rmax}{\pi\ell/2}
\newcommand{\mpp}{M_{\rm PP}}
\begin{document}

\title{\bf Gravitational collapse in 2+1 dimensional AdS spacetime }

\author{
Frans Pretorius$^1$\footnote{Electronic address: \tt fransp@physics.ubc.ca\hfil}
\ and
Matthew W. Choptuik$^{1,2,3}$
\footnote{Electronic address: \tt choptuik@physics.ubc.ca\hfil} 
\\
\\
\\
$^1$ Department of Physics and Astronomy, University of British Columbia, \\
Vancouver BC, V6T 1Z1   Canada \\
$^2$ CIAR Gravity and Cosmology Program \\
$^3$ Center for Relativity, University of Texas at Austin, TX 78712-1081 USA
\\
}

\maketitle

\begin{abstract} 
We present results of numerical simulations of the formation
of black holes from the gravitational collapse of a
massless, minimally-coupled scalar field in 2+1 dimensional,
axially-symmetric,
anti de-Sitter (AdS) spacetime. The geometry exterior to the
event horizon approaches the BTZ solution, showing no evidence
of scalar `hair'. To study the interior structure we implement
a variant of black-hole excision, which we call
singularity excision. We find that interior to the 
event horizon a strong, spacelike curvature singularity develops. 
We study the critical behavior at the threshold of 
black hole formation, and find a continuously self-similar solution
and corresponding mass-scaling exponent of approximately 1.2. The
critical solution is universal to within a phase that is
related to the angle deficit of the spacetime. 
\end{abstract}

\section{Introduction}

The past several years has seen growing interest in the properties and 
dynamics of asymptotically anti de-Sitter (AdS) spacetimes, predominantly 
due to the discovery of black hole solutions in 2+1 dimensional AdS spacetime 
\cite{BTZ} and the AdS/CFT conjecture \cite{ads_cft}. The existence of 
vacuum \footnote{with a negative cosmological constant} black holes 
(also called BTZ black holes) is surprising because the local solution to the 
field equations is isometric to AdS, and hence has constant curvature. 
What makes a BTZ spacetime different from AdS is its global structure, 
which can be obtained by making appropriate identifications within the 
universal covering space of AdS \cite{BHTZ}. The natural question that such a 
construction poses is: how similar are these black holes to their more 
familiar 3+1 dimensional (4D) counter-parts? In particular, do these
black holes have 
thermodynamic properties when considered within the framework of a 
quantum theory, and can they form through dynamical 
collapse processes? It turns out that BTZ black holes {\em do} bear
striking resemblance to 4D black holes in many respects 
(see \cite{review} for review articles). In this paper we present the 
results of a numerical study of the collapse and formation of 
non-rotating BTZ black holes from a massless scalar field in 2+1D 
AdS spacetime. Of particular interest is whether critical phenomena
\cite{choptuik}
are present at the {\em threshold} of black hole formation---namely 
if by fine-tuning of initial data, we can make the system 
asymptote (at ``intermediate times'') to a solution
which is universal in the sense of not depending on details of the initial 
data.  Furthermore, if the black hole transition is ``Type II'', so 
that there is {\em no} smallest mass of black hole which can be formed,
then we expect their to be a scaling relation
for the black hole mass of the form $M=K (p-p^\star) ^{2\gamma}$. Here
$p$ is a parameter in a family of initial data such that 
$p=p^\star$ is the critical solution, $K$ is a family dependent constant
and $\gamma$ is a universal exponent
(see \cite{cp_review} for a recent review). The `extra' factor of
2 in the exponent is expected for BTZ black holes---see section \ref{sec_crit}.
As we will show, it turns out that the system {\em does} exhibit a continuously 
self-similar (CSS) solution in the critical limit, with a scaling exponent
$\gamma=1.2 \pm 0.05$. \par 
Earlier works on black hole formation in AdS 
considered disks of dust \cite{dust}, null radiation \cite{null_rad},
thin dust rings \cite{shell}, and the collision of 
point-particles \cite{particles}. In the case of dust-ring collapse
Peleg and Steif found a scaling exponent of $1/2$ at the transition between
black hole and naked singularity formation. Birmingham and Sen found the
same exponent at the threshold of formation in the case of 
colliding particles. 
Husain and Olivier have also
studied the massless scalar field in 2+1 dimensions using a double
null formalism, and have formed black holes with their code
\cite{dnull}.\par

Our paper is organized as follows. In section \ref{setup} we describe 
the system of coordinates and numerical scheme we have chosen to use, 
and the resultant field equations and boundary conditions. An interesting
consequence of our analysis 
is that we are unable to derive boundary conditions for the scalar field
at the edge of the universe that are analogous to the out-going radiation 
conditions often employed in numerical relativity. In AdS spacetime the
scalar field reaches time-like infinity $\Scri$ in 
{\em finite} proper time as measured by a central observer, and the only
consistent boundary conditions we can place on the scalar field 
confine it to the universe. This is reassuring 
from the standpoint of global energy conservation, but
complicates the search for the universal scaling relation between black hole 
mass and parameter-space distance to the critical solution.
The system behaves as if the scalar field is within a finite sized box, and so 
when a black hole forms
all of the scalar field initially present eventually falls into the hole.
$M(p)$ is therefore trivially a function of how the initial energy 
distribution scales with $p$. \par
In section \ref{results} we present results from the evolution of
several families of initial data, focusing on critical behavior.
To obtain $\gamma$, we follow the work of Garfinkle and Duncan\cite {garfinkle},
and examine the scaling of the maximum
value attained by the curvature scalar $R$ in the sub-critical regime.
We also study the effect that a central point particle (characterized
by the angle deficit of the spacetime) has on the critical solution. 
As expected, we find that the more massive the point particle, 
the smaller the initial
amplitude of the scalar field that gives rise to the critical solution. 
One might thus expect to have a one-parameter family of critical solutions with 
an overall scale related to the particle mass. It {\em is} surprising,
therefore, that the scalar field always grows to the {\em same}
amplitude in a near-critical evolution. A phase shift in 
central proper time
is the only qualitative difference attributable to the mass 
of the particle.
At the end of section \ref{results} we study the interior structure of 
black holes that form, giving evidence 
that a `crushing' spacelike curvature singularity forms within the event
horizon.
Thus the interior structure is significantly different from the BTZ solution, 
which has constant curvature (though the BTZ singularity is still crushing for 
extended objects falling into it). 

\section{The Einstein Klein-Gordon system in AdS spacetime}\label{setup}

We solve the Einstein field equations in 3 spacetime dimensions with 
cosmological constant $\Lambda \equiv -1/\ell^2$, coupled to a 
massless Klein-Gordon (KG) field
\begin{equation}\label{ekg}
R_{ab}-\frac{1}{2}R g_{ab} + \Lambda g_{ab} = \kappa T_{ab},
\end{equation}
where the stress-energy-momentum tensor for the KG field $\phi$ 
is \cite{wald}
\begin{equation}\label{kgset}
T_{ab}=\phi_{;a}\phi_{;b} - \frac{1}{2} g_{ab} \phi_{;c} 
\phi^{;c}.
\end{equation}
Covariant differentiation is denoted by a semi-colon, while
a comma denotes partial differentiation. 
We only consider circularly symmetric configurations of a 
minimally-coupled scalar field in this paper. Hence, 
$\phi$ satisfies the wave equation
\begin{equation}\label{wave}
\Box \phi = \phi_{;a}^{\ \ a} = 0,
\end{equation}
and in coordinates ($t,r,\theta$) adapted to the symmetry, characterized by 
the Killing vector $\partial/\partial \theta$, $\phi(r,t)$ is only 
a function of the radial coordinate, $r$, and time coordinate, $t$.

One of the many peculiar features of AdS spacetime is its causal 
structure. In particular, null infinity $\Scri$ is time-like, and 
any observer living in AdS spacetime can send and receive light-like 
signals to and from $\Scri$ in finite proper time 
\cite{hawking_ellis}. These properties of AdS make it 
challenging to deal with numerically, as the scalar field 
traverses the entire universe on a local dynamical time-scale. 
Also, as we will show in section \ref{bcs}, the only regular 
boundary conditions on the field $\phi$ at $\Scri$ are Dirichlet 
conditions, so we cannot ignore the unusual causal structure of 
the spacetime by, for instance, placing out-going radiation 
boundary conditions on $\phi$ at a finite proper distance 
from the origin. For these reasons, we adopt a coordinate system
in which the metric takes the form:
\begin{equation}\label{metric}
ds^2=\frac{e^{2A(r,t)}}{\cos^2(r/\ell)} \left(dr^2-dt^2\right) + 
\ell^2 \tan^2(r/\ell) e^{2B(r,t)}d\theta^2. 
\end{equation}
$A(r,t)$ and $B(r,t)$ are arbitrary functions of $(r,t)$, and it is 
straight-forward to show that when $A=B=0$ the above metric 
describes AdS spacetime; i.e. it is a solution to (\ref{ekg}) 
with $T_{ab}=0$. Notice that, in this metric, radial null geodesics 
travel with constant coordinate speed $dr/dt = \pm 1$, 
and $\Scri$ is at $r=\pi \ell/2$. The metric is singular at $\Scri$, 
but we can place regular boundary conditions on $A$ and $B$ there, so 
that the spacetime is asymptotically AdS. 
Also, if we interpret $\theta$ as a periodic 
angular variable then the above metric has the correct 
topology to represent a BTZ black hole, as the topological 
censorship theorems require that the boundary at infinity share 
the topology of any event horizon that may exist in the interior 
of the spacetime \cite{topcen}. However, for the non-rotating collapse 
described in this paper, $\theta$ has no dynamical significance.  

Defining
\begin{equation}\label{PHI_PI}
\Phi(r,t)=\phi_{,r} \ , \ \ \ \Pi(r,t)=\phi_{,t}
\end{equation}
and using units where $\kappa=4 \pi$, we get the following set of 
equations upon  expanding (\ref{ekg})--(\ref{wave}) with the 
metric (\ref{metric}):
\begin{equation}\label{A}
A_{,rr} - A_{,tt} + \frac{(1-e^{2A})}{\ell^2 \cos^2(r/\ell)} + 
2\pi (\Phi^2-\Pi^2) =0,
\end{equation}
\begin{equation}\label{B}
B_{,rr} - B_{,tt} + B_{,r} \left(B_{,r} + \frac{2}{\ell 
\cos(r/\ell)\sin(r/\ell)} \right) -(B_{,t})^2 + \frac{2(1-
e^{2A})}{\ell^2 cos^2(r/\ell)} = 0,
\end{equation}
\begin{eqnarray}\label{hamil}
B_{,rr} + B_{,r} \left( B_{,r} - A_{,r} + 
\frac{1+\cos^2(r/\ell)}{\ell \cos(r/\ell)\sin(r/\ell)}\right) - \nonumber \\ 
\frac{A_{,r}}{\ell\cos(r/\ell)\sin(r/\ell)} - A_{,t}B_{,t} + 
\frac{(1-e^{2A})}{\ell^2 cos^2(r/\ell)} + 2\pi (\Phi^2 + \Pi^2) 
=0,
\end{eqnarray}
\begin{equation}\label{momen}
B_{,rt} + B_{,t}\left(B_{,r}-A_{,r}+\frac{cot(r/\ell)}{\ell} 
\right) - A_{,t} \left( B_{,r} + 
\frac{1}{\ell\sin(r/\ell)\cos(r/\ell)}\right) + 4 \pi \Phi \Pi = 
0
\end{equation}
and
\begin{equation}\label{wave_exp}
\left[\tan(r/\ell) e^{B} \Phi\right]_{,r} - \tan(r/\ell) 
\left[e^{B} \Pi \right]_{t} =0.
\end{equation}

Within the context of the 3+1, or ADM, formalism,
equations (\ref{hamil}) and (\ref{momen}) 
are the Hamiltonian and momentum constraints respectively, while 
equations (\ref{A}) and (\ref{B}) are combinations of the 
evolution and constraint equations. Equation (\ref{wave_exp}) is 
the wave equation for the scalar field. There are two unknown 
geometric variables---$A(r,t)$ and $B(r,t)$; hence one needs to use at least 
two of the four  equations (\ref{A}) - (\ref{momen}) to dynamically 
determine the geometry. 
In this work, we have chosen to use equations (\ref{A}) and (\ref{B})
to update $A$ and $B$.  As is common practice in such a ``free evolution
scheme'', we can then use 
residuals of the constraints (\ref{hamil}) and (\ref{momen}) as one way
of estimating the level of error in our solution.

With regards to initial conditions, 
we choose to freely specify $\Phi(r,0)$ and $\Pi(r,0)$ (we {\em have} to specify
two scalar-field degrees of freedom at each $r$), as well as $B(r,0)$ and 
$B_{,t}(r,0)$. $A(r,0)$ and $A_{,t}(r,0)$ are then fixed from the constraint 
equations (see sec. \ref{sec_ic} for more details).  This procedure is 
clearly somewhat {\em ad hoc}, but has worked very well in our study.

The Ricci scalar of this spacetime is
\begin{equation}\label{rs}
R=\frac{4\pi \cos(r/\ell)^2}{e^{2A} \ell^2} \left(\Phi^2-\Pi^2\right)
- \frac{6}{\ell^2}.
\end{equation}
The Weyl tensor is zero, and other non-zero curvature scalars can be 
expressed as polynomial functions of $R$.

\subsection {Regularity conditions} \label{bcs}

We require that the solution for our dynamical variables $A(r,t), B(r,t), 
\Phi(r,t)$ and $\Pi(r,t)$ be regular at the origin, $r=0$, and at $\Scri$, 
$r=\pi \ell/2$. The field equations then essentially dictate the 
allowed boundary conditions on these variables. By inspection of 
(\ref{A})--(\ref{wave_exp}) we obtain the following conditions.
At $r=0$
\begin{eqnarray}
A_{,t}(0,t)&=&B_{,t}(0,t) \label{AB_bcmin}\\ 
A_{,r}(0,t)&=&0 \label{A_bcmin}\\
B_{,r}(0,t)&=&0 \label{B_bcmin}\\
\Phi(0,t)&=&0 \label{phi_bcmin}\\
\Pi_{,r}(0,t)&=&0 \label{pi_bcmin}
\end{eqnarray}
and at $r=\pi\ell/2$
\begin{eqnarray}
A(\pi\ell/2,t)&=&A_{,r}(\pi\ell/2,t)=A_{,t}(\pi\ell/2,t)=0 \label{A_bcmax}\\
B_{,r}(\pi\ell/2,t)&=&0 \\
\Phi(\pi\ell/2,t)&=&0 \\
\Pi(\pi\ell/2,t)&=&0 \label{Pi_bcmax}. 
\end{eqnarray}
Note that condition (\ref{pi_bcmin}) on $\Pi(0,t)$ is a direct consequence of
the defining relation for $\Pi(r,t)$~(\ref{PHI_PI}), and the regularity 
condition for $\Phi(0,t)$ (\ref{phi_bcmin}).
Also note that we have multiple conditions for $B$ at the outer 
boundary, and for $A$ and $B$ at the origin. We have chosen to 
implement the Neumann conditions for $A$ and $B$ at the origin 
and the Dirichlet condition for $A$ at $\Scri$, and then to monitor 
the other conditions as a consistency check during evolution. 
Conditions (\ref{A_bcmax})--({\ref{Pi_bcmax}) ensure that the 
spacetime is asymptotically AdS.

It is interesting that the field equations enforce Dirichlet 
boundary conditions on $\Phi$ and $\Pi$, effectively preventing 
us from implementing out-going radiation boundary conditions at 
$\Scri$ (if we wanted to let the field ``leak out of the universe''
when it reaches $\Scri$). To see this more clearly, consider the 
energy fluxes $T_{ab}\eta^a \eta^b$ and $T_{ab} \ell^a \ell^b$ 
along outgoing and ingoing null vectors, $\ell^a$ and $\eta^a$, respectively,
normalized so that $\ell^a \eta_a = -1$
\begin{equation}\label{null_out}
\ell^a = \frac{\cos(r/\ell)}{\sqrt{2}e^A} \left[
\frac{\partial}{\partial t} + \frac{\partial}{\partial r} 
\right]^{a}
\end{equation}
\begin{equation}\label{null_in}
\eta^a = \frac{\cos(r/\ell)}{\sqrt{2}e^A} \left[
\frac{\partial}{\partial t} - \frac{\partial}{\partial r} 
\right]^{a}
\end{equation}
A straight forward calculation using (\ref{kgset}) gives 
\begin{equation}\label{null_flux}
E_\pm=\frac{\cos(r/\ell)^2 (\Phi \pm \Pi)^2}{2 e^{2A}}, 
\end{equation}
where $E_+$ is the influx and $E_-$ the outflux. Thus no-
outflux/influx boundary conditions can be obtained in the usual 
way by differentiating $\Phi \pm \Pi$ with respect to $r$ and $t$ 
in turn, and utilizing the fact that, from (\ref{PHI_PI}), $\Phi_{,t}=\Pi_{,r}$:
\begin{eqnarray} \label{outrad1}
\Phi_{,r} \pm \Phi_{,t} = 0 \\ \label{outrad2}
\Pi_{,r} \pm \Pi_{,t} = 0 \, . \\ 
\end{eqnarray}
Here, the plus sign corresponds to no-influx, and the minus sign 
to no-outflux. However, at the outer boundary, regularity forces 
$\Phi(\rmax,t)=\Pi(\rmax,t)=0$, and hence 
$\Phi_{,t}(\rmax,t)=\Pi_{,t}(\rmax,t)=0$, so there is no 
distinction between the no-influx and no-outflux condition. The 
only situation consistent with both conditions is that {\em no}
flux crosses the outer boundary in either direction. 
Even when we try to derive no-outflux/influx conditions with the asymptotic 
behavior of $\phi$ factored out, namely letting 
$\phi=\cos^2(r/\ell)\hat{\phi}$ and placing boundary conditions 
on $\hat{\phi}$, we find that the wave equation on $\Scri$ cannot 
distinguish between no-outflux and no-influx conditions.
Also, in early experiments we were unable to obtain stable numerical 
evolution with the no-influx boundary conditions (\ref{outrad1}) and
(\ref{outrad2}) applied at a finite proper circumference,
corresponding to $r<\pi\ell/2$}.  
The Dirichlet boundary condition at $\Scri$ is also consistent with 
the behavior of a massive scalar field in an AdS background, 
where an infinite effective-potential barrier prevents any of the 
field from reaching $\Scri$, regardless of how small the mass is.
Of course, all of this does not mean that an effective outgoing 
radiation condition can not be implemented for the massless
field in asympotically AdS spacetimes. In any case, in the context of the
current study, we would be apt to view such a condition as a numerical 
convenience, rather than being of any intrinsic physical interest.

\subsection{Initial conditions}\label{sec_ic}

For initial conditions at $t=0$, we are free to specify the scalar field 
gradients
$\Phi(r,0)$ and $\Pi(r,0)$, the metric function $B(r,0)$ and its time derivative
$B_{,t}(r,0)$. We then numerically solve for $A(r,0)$ and $A_{,t}(r,0)$ using 
the hamiltonian and momentum constraints (\ref{hamil}) and ({\ref{momen}).
The freedom that 
we have to specify $B(r,0)$ amounts to a choice of what the proper
circumference, ($\ell \tan(r/\ell) e^B$), and its initial time derivative
are, as a function of the radial coordinate $r$. That we do not have the 
freedom to choose $B$ for all time
is a consequence of the gauge condition that radial light-like signals
travel with unit coordinate velocity. 
For simplicity we set $B(r,0)=B_{,t}(r,0)=0$.

We believe (though are unable to prove so), that the set of conditions just
described is capable of generating all possible initial data,
which is regular and free of trapped surfaces,
for the minimally-coupled scalar field in asymptotically AdS
spacetime (in 2+1 dimensions). The presence of trapped surfaces
at $t=0$ is incompatible with the conditions on $B(r,0)$ and
$B_{,t}(r,0)$--- 
in our coordinate system $dr/dt=1$
along an outgoing null curve, and hence a non-zero $B(r,0)$ and/or $B_{,t}(r,0)$ is 
required to describe non-positive outward null-expansion.
However, in this study we are only interested in initial data 
that {\em is} free of trapped surfaces, so the conditions on 
$B(r,0)$ and $B_{,t}(r,0)$ are not restrictive. 

For the initial scalar field profile, $\phi(r,0)$, we consider three 
families of functions---a gaussian curve raised to the $n^{th}$ power
\begin{equation}\label{gauss}
\phi(r,0)= \amp e^{\left((r-\ro)/\width \right)^{2n}},
\end{equation}
a `kink' (based on an {$\arctan$} function) 
for which $\Phi=\partial\phi/\partial r$ is
\begin{equation}\label{kink}
\Phi(r,0)= \frac{-2 \amp \sqrt{\width} \cos(r/\ell)\sin(r/\ell) 
\left[-\ell\sin(r/\ell)\cos(r/\ell)+2(r-\ro)(1-2\sin(r/\ell)^2)
\right]
e^{-(r-\ro)^2/\width^2}}
{\pi\ell\left[\width\sin(r/\ell)^4\cos(r/\ell)^4+(r-\ro)^2\right]} 
\end{equation}
and a family of harmonic functions \footnote{we call these functions 
`harmonic' because without back-reaction and for initially
static configurations ($\Pi(r,0)=0$) the exact solution to the
wave equation is periodic in time.}
\begin{equation}\label{harm}
\phi(r,0)= \amp \cos^2(r n/\ell),
\end{equation}
where $\amp,\ro,\width$ and $\cosn$ are constant parameters.
Then, depending upon whether we want to model 
initially ingoing, outgoing or static 
fields, we set $\Pi(r,0)=\Phi(r,0)$, $\Pi(r,0)=-\Phi(r,0)$ or $\Pi(r,0)=0$ 
respectively. Note that this method cannot give purely ingoing
or outgoing pulses---$\Pi(r,t)=\pm\Phi(r,t)$ is {\em not} 
an exact solution to the wave 
equation, and a little bit of energy always propagates in the
opposite direction to that desired. 

As noted previously, we set $B(r,0)=B_{,t}(r,0)=0$.  The remaining geometric
variables, $A(r,0)$ and $A_{,t}(r,0)$ are then computed
from the Hamiltonian and momentum constraints 
(\ref{hamil}) and (\ref{momen}). We integrate the constraints outwards from
$r=0$, setting $A_{,t}(0,0)=0$.
For the most part we will consider the collapse of a scalar
field initially exterior to empty AdS space. This corresponds 
to setting $A(0,0)=0$. However, in section \ref{sec_pp},
we will briefly consider the 
effect of collapsing the field in the presence of a point particle at
the origin, the calculation of which involves introducing 
an angle deficit into the spacetime.
From the metric (or by examining the parallel 
transport of a vector about $r=0$ in 
an infinitesimal loop), the angle deficit 
$\omega$ at $t=0$ is related to $A(0,0)$
as follows:
\begin{equation}\label{deficit}
\omega=2\pi(1-e^{A(0,0)}).
\end{equation}
Of more interest is the relationship between $A(0,0)$ and the mass of 
the point-particle, $M_{pp}$. The remainder of this section is
devoted to finding this relationship, and in the process we 
will define a general mass aspect function $M(r,t)$ for the spacetime.

When the scalar field gradients identically vanish (which they
do at $\Scri$, and, to an excellent approximation, at $r=0$ for
the initial data that we consider), the Hamiltonian 
constraint has the simple solution
\begin{equation}\label{A_phi0}
e^{2A}=\frac{k}{k-\cos^2(r/\ell)},
\end{equation}
where $k$ is a constant of integration. We can relate $k$ to the BTZ 
mass parameter $M$ of the spacetime by appealing to the
usual form in which the BTZ solution is expressed:
\begin{equation}\label{metric_BTZ}
ds^2=-(-M+\rb^2/\ell^2) d\tb^2 + \frac{1}{-M+\rb^2/\ell^2} 
d\rb^2 + \rb^2 d\theta^2.
\end{equation}
$M=-1$ is AdS spacetime, $M \ge 0$ are black hole solutions and 
$M<0, M\neq -1$ are spacetimes with conical singularities, or point particles
at the origin 
(the range of $\rb$ is from $0$ to $\infty$). 
For general (non-vacuum)
solutions let us define the mass aspect $M(\rb,\tb)$ as follows
\begin{equation}\label{Mdef}
|\nabla \rb |^2 \equiv -M(\rb,\tb) + \rb^2/\ell^2. 
\end{equation}
Then, in our coordinate system (\ref{metric}), $M(r,t)$ takes 
the following form 
\begin{equation}\label{M}
M(r,t)=e^{2(B-A)}\left[e^{2A}\tan^2(r/\ell) + 
\ell^2 \sin^2(r/\ell)((B_{,t})^2-(B_{,r})^2) 
- 2 \ell \tan(r/\ell) B_{,r} - \sec^2(r/\ell) \right].
\end{equation}   
Using the field equations (\ref{A})-(\ref{momen}) it is straightforward 
to show that $M$ is a conserved quantity in regions of the spacetime 
where $\Phi$ and $\Pi$ are zero (in particular at $\Scri$). At $t=0$, 
where $B=B_{,t}=0$, we can substitute (\ref{A_phi0}) into (\ref{M}) to 
find $k$: 
\begin{equation}\label{mbtz} 
k=\frac{1}{1+M}.
\end{equation}
When $M \ge 0$ our metric (\ref{metric}) with the chosen initial conditions
is singular at the horizon of 
an empty BTZ spacetime, but the metric turns out to be well
behaved at $t=0$ for initial data that does not contain trapped surfaces 
\footnote{however, because of our choice of gauge, we 
know that the coordinate system must become singular 
within one light-crossing time (LCT) of the formation of a 
black hole. The event horizon is a null hypersurface travelling outward with 
unit coordinate velocity, thus the coordinate distance between the event
horizon and $\Scri$ will go to zero within a time $t=\pi\ell/2$.}. 

Finally, from (\ref{A_phi0}) and (\ref{mbtz}) the contribution, $\mpp$, of 
the point particle to the mass of the spacetime is
\begin{equation}\label{m_pp}
\mpp=1-e^{-2A(0,0)}
\end{equation}

\subsection{Numerical Scheme}

We solve the set of equations (\ref{A}),(\ref{B}) and (\ref{wave_exp}) by 
converting them to a system of finite difference equations on a 
uniform coordinate grid using
a two-time level Crank-Nicholson scheme. We also add Kreiss-Oliger style
dissipation \cite{KO} to control high-frequency solution components; this is 
crucial for the stability of our method.

At first, we used standard
2nd order accurate 3-point finite difference stencils for the spatial
derivatives at each time level---centered-difference operators at
interior points, a forward-difference operator at the inner
boundary and a backward-difference operator at the outer boundary.
However, we found that these operators excited a small instability
in the metric variables in the vicinity of the outer boundary. 
The resultant ripples would 
propagate inwards and cause problems in situations where black hole
formation was imminent. 
The primary source of these ripples was truncation error in the solution
$A$ exciting small oscillations in $B$. Specifically, $A$ acts as a source term
in the evolution equation for $B$ (\ref{B}), and $B$ happens to be
very sensitive to small errors in $A$ near the outer boundary 
(essentially since the leading term of $A$, when considered as a power
series in $\cos^2(r/\ell)$, cancels with the spatial derivatives of $B$
in (\ref{B}) initially, and so higher order, less accurately known terms
of $A$ are responsible for $B$'s ``acceleration''). 
To reduce these problems we now use a 5-point, 4th order 
accurate spatial derivative operator at interior grid points, and 6-point
4th order backward and forward operators 
near boundaries that have the same truncation
error as the interior operator.  Also, we find that using the momentum
constraint (\ref{momen}) to solve 
for $A$ at the next to last grid point is necessary 
to obtain convergence of the solution as we go to finer
spatial resolution (for some as yet unknown reason the evolution equation
was exciting a growing mode on finer grids at that point). 
The program to perform the evolution was written in Fortran 77 and
RNPL (Rapid Numerical Prototyping Language \cite{rnpl}); 
animations and pictures from several evolutions can be obtained
from our website \cite{web}.

\subsection{Detecting black holes and excising singularities}\label{detect_bhs}

To detect black hole formation we search for trapped surfaces, 
defined to be surfaces where the expansion of outgoing null
curves normal to the surface is negative. If
cosmic censorship holds, then trapped surfaces are always found within the
event horizon of a black hole, though at the end of the simulation
we can trace null rays backwards from $\Scri$ to confirm this. 
In our coordinate system the condition for a surface to be trapped is
\begin{equation}\label{trapped}
1+\ell\cos(r/\ell)\sin(r/\ell)(B_{,r}+B_{,t}) < 0.
\end{equation}
We estimate the mass of the black hole by monitoring the proper
circumference $2\pi \ell\tan(r_{{\rm AH}}/\ell)e^{B(r_{{\rm AH}},t)}$ of 
the apparent horizon (the outer-most trapped 
surface), and use the relationship between BTZ black hole mass
and event horizon circumference ((\ref{metric_BTZ})---the horizon
is at $\rb=\sqrt{M} \ell$)):
\begin{equation}\label{m_approx}
M \approx \tan^2(r_{{\rm AH}}/\ell)e^{2B(r_{{\rm AH}},t)}.
\end{equation}
If all of the scalar field is absorbed by the black hole during
evolution,
then the estimated mass should eventually become equal to the initial, 
asymptotic mass of the spacetime as given by (\ref{M}) in the limit
$r\rightarrow\pi\ell/2$.

As we will show in section \ref{sing}, shortly after an apparent horizon(AH)
forms, we find what appears to be a spacelike curvature singularity
forming within the AH.  If we use a straightforward evolution scheme, the 
metric and scalar field variables quickly diverge, and any given simulation 
just as quickly breaks down.  At the same time, we would 
like to probe the structure of the spacetime approaching the
singularity, as well as to continue to following the evolution outside the AH 
as long as our coordinate system allows (approximately 1 light-crossing time).
To accomplish this, we have implemented {\em singularity excision}, a technique 
fundamentally motivated by
the black-hole-excision strategy first proposed by Unruh~\cite{ah}. \par
Our excision strategy is as follows.
We monitor the magnitude of the metric
variables, and when they grow beyond a certain 
threshold \footnote{for a threshold we choose 
a number that is sufficiently large so we
are fairly certain (from past experiments) that if any variable grows
beyond the threshold then a crash is imminent} at
any point we excise that point plus a small buffer zone (of 4 to 6 grid points)
on either side of it. (Note that the non-excised 
region of the grid will no longer be
contiguous if the excised point is further away from the original grid
boundaries than the size of the buffer zone).
At the new grid boundaries exterior to the
excised region, we continue to solve for the metric and field variables
using the evolution equations, but replace all centered-difference
operators with forward and backward-difference operators, as appropriate,
so that the solution is not ``numerically influenced'' 
by the excised grid points.
Physically, the solution that one would obtain within the causal future
of the excised zone is meaningless, so we also remove this region
of the grid during subsequent evolution.
In our coordinate system this is easy to implement,
as radial null curves travel at constant, unit coordinate velocity. 
Thus, if our grid-spacing is $\Delta r$, 
after an amount of time $\Delta t=\Delta r$ 
we expand the excised region by 1 grid point on either side. Also, we 
continue to monitor the metric variables on the remainder of the grid, and
when they grow beyond the threshold at any 
other points we expand the excised region 
to include those points (and a buffer). Thus the excised piece of the
grid is always contiguous. In principle, it 
would not be difficult to keep track
of multiple excised zones, though we did not find it necessary to
do so for the 
interior solution shown in sec. \ref{sing}---a single zone is sufficient 
to obtain a good view of all of the interior up to the 
putative spacetime singularity.\par
We have tested the singularity excision scheme by excising a light cone 
from a solution that remains regular, and verifying that the excised solution 
{\em does} converge to the regular solution as $\Delta r$ decreases. \par
In summary, we briefly clarify the difference between 
singularity and black hole excision.  First, notice
that we never use trapped surfaces to trigger the excision of a region of the
grid. 
Thus, our code could, without modification, excise naked and
coordinate singularities. The boundary of the excised region is always
null or spacelike, so the scheme might not be able to distinguish
between timelike and null curvature singularities. However, if a 
timelike singularity was encountered, it may still be
possible to deduce its nature by examining the curvature
invariants just exterior to the excised surface. For example,
suppose during evolution a light-like region was excised, and
curvature invariants started diverging as one approached
the initial excised point, yet remained relatively ``small'' and
finite just outside the future light-cone of the excised point, 
then one would have reasonable evidence for a timelike singularity.
Second, with the singularity excision scheme, we 
excise only the region of the grid to the causal future of the singularity.
In the
case of a black hole spacetime, this results in
a more complete view of the spacetime than what one would
obtain with the standard black hole excision strategy (which
would have in Fig. \ref{RS133051_bw}, for example, excised
the region of the spacetime labeled ``region of trapped
surfaces'', and everything to the left of it).

\section{Results}\label{results}

In this section we discuss results from the evolution of several
sets of initial data, focusing on the threshold of black hole formation. 
For convenience
we set $\ell=2/\pi$ so that $\Scri$ is at $r=1$, 
though the results presented here are valid
for any non-zero, finite $\ell$, through an appropriate rescaling of the 
metric variables and scalar field gradients. Specifically, 
consider the following coordinate transformation 
\begin{equation}\label{l_scale_trans}
\tilde{r}=\frac{r}{\ell}, \ \ \ \tilde{t}=\frac{t}{\ell},
\end{equation}
with $\tilde{r}$ defined on the range $[0,\pi/2]$.
Then it is easy to see that the $\ell$ dependence cancels from
all equations (\ref{A})-(\ref{wave_exp}) when expressed in terms
of $\tilde{r}$ and $\tilde{t}$. So, given a solution
$A(\tilde{r},\tilde{t})$, $B(\tilde{r},\tilde{t})$,
$\Phi(\tilde{r},\tilde{t})$ and $\Pi(\tilde{r},\tilde{t})$
to the rescaled field equations
we can find a corresponding solution for any $\ell$ 
by inverting the transformation
(\ref{l_scale_trans}) (see also (\ref{PHI_PI})):
\begin{eqnarray}\label{l_scale_vars}
A(\tilde{r},\tilde{t}) &\rightarrow& A(r/\ell,t/\ell) \\ \nonumber
B(\tilde{r},\tilde{t}) &\rightarrow& B(r/\ell,t/\ell) \\ \nonumber
\Phi(\tilde{r},\tilde{t}) &\rightarrow& \ell \Phi(r/\ell,t/\ell) \\ \nonumber
\Pi(\tilde{r},\tilde{t}) &\rightarrow& \ell \Pi(r/\ell,t/\ell),
\end{eqnarray}
with $r$ ranging from $0$ to $\pi\ell/2$. Notice that the initial 
energy density, being proportional to ($\Phi^2+\Pi^2$), scales like $\ell^2$,
so there is no straight-forward method to extrapolate a solution to the 
limit of zero cosmological constant, where $\ell \rightarrow \infty$.

We present results from 4 families of initial data: an ingoing gaussian 
((\ref{gauss}) with $n=1$), an ingoing squared gaussian ((\ref{gauss}) 
with $n=2$),
an ingoing kink (\ref{kink}), and a time-symmetric, $n=1$, 
harmonic function (\ref{harm}).
In each case we vary the amplitude $\amp$ when tuning to the black hole 
threshold\footnote{though we did check (for the gaussian) that we get
the same critical solution when tuning the width, keeping the amplitude fixed},
and for the first three families we have chosen $\width=0.05$ and
$\ro=0.2$.  
Except in section \ref{sec_pp}, where we briefly study collapse onto a 
point-particle, we have set $A(0,0)=0$ in all cases, corresponding
to angle deficit-free spacetimes. 
The 3 ingoing families were simulated using a finest numerical 
grid of size 4096 points, with a Courant
factor of 0.1; thus 40960 time steps are required per light-crossing time (for
some of the critical solutions presented in the next section a grid size of
8192 points was used with a Courant factor of 0.2)
For the time-symmetric $\cos^2$ function we do not need as many points to
get good convergence results (because of the milder field gradients), 
so that the highest resolution required for that family was a 1024-point grid. 
In fact, we get 
acceptable results even after 50 light-crossing times with 1024 points
for the $\cos^2$ data, whereas the more compact ingoing families start
having noticeable errors (estimated from convergence tests) in near-critical 
evolution after 3-4 LCT's with 4096 points. 

Fig. \ref{phi_t0} shows the initial scalar field gradient, $\Phi(r,0)$, of 
typical amplitude for each of the families.  Fig. \ref{AB_t0_t6} shows
the metric function $A(r,0)$ for a gaussian (the other families have
similar shapes for $A$), and for later reference we show how $A(r,t)$ and 
$B(r,t)$ have evolved at $t=0.6$.
In order to provide the reader with some feeling for the dynamics
of a ``typical'' evolution, 
Fig. \ref{g1_phi_1302_bw} shows a ``space-time'' plot 
of the evolution of a sample gaussian with
$\amp=0.1302$ that does not form a black hole within 4 LCT's (and it should not,
as the asymptotic mass of the spacetime is $-1.062 \mbox{x} 10^{-2}$).

\begin{figure}
\epsfxsize=17cm
\centerline{\epsffile{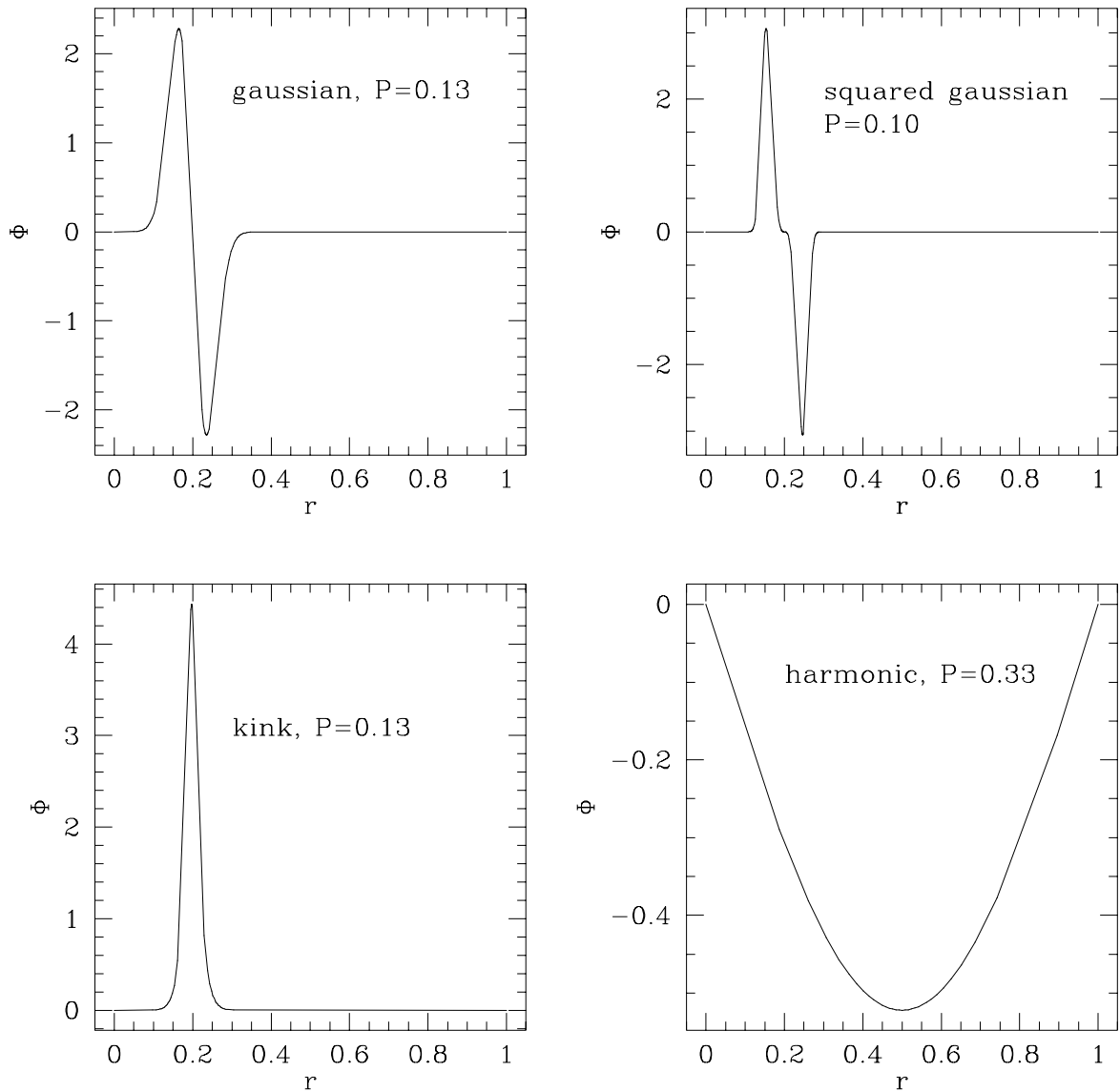}}
\caption{\label{phi_t0}
$\Phi(r,0) = \phi_{,r}(r,0)$ for
each family of initial data studied. The three compact families
are initially ingoing, thus $\Pi(r,0)=\Phi(r,0)$, while the harmonic function is
time-symmetric with $\Pi(r,0)=0$ ($\ell=2/\pi$, so $\Scri$ is at $r=1$).
}
\end{figure}

\begin{figure}
\epsfxsize=17cm
\centerline{\epsffile{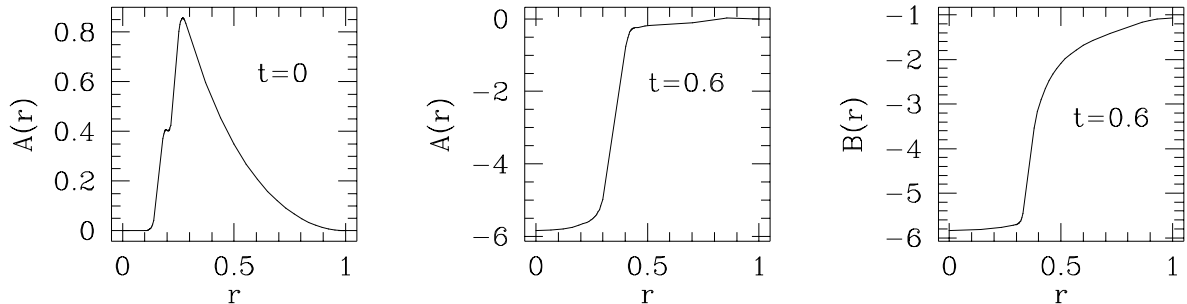}}
\caption{\label{AB_t0_t6}
$A(r,0)$ (left-most figure) for a gaussian with $\amp=0.133051$, 
as obtained by solving the
Hamiltonian constraint with $B(r,0)=0$. This amplitude is used as an example in
section \ref{sing} when we discuss the singularity structure, so for reference
we also show $A(r,0.6)$ and $B(r,0.6)$. Notice in particular how large and
negative $B$ is towards the origin, indicating that in this region of the grid
we are looking at very small scales in the problem (the proper circumference
element is $\rb=\ell\tan(r/\ell) e^B$).
}
\end{figure}

\begin{figure}
\epsfxsize=10cm
\centerline{\epsffile{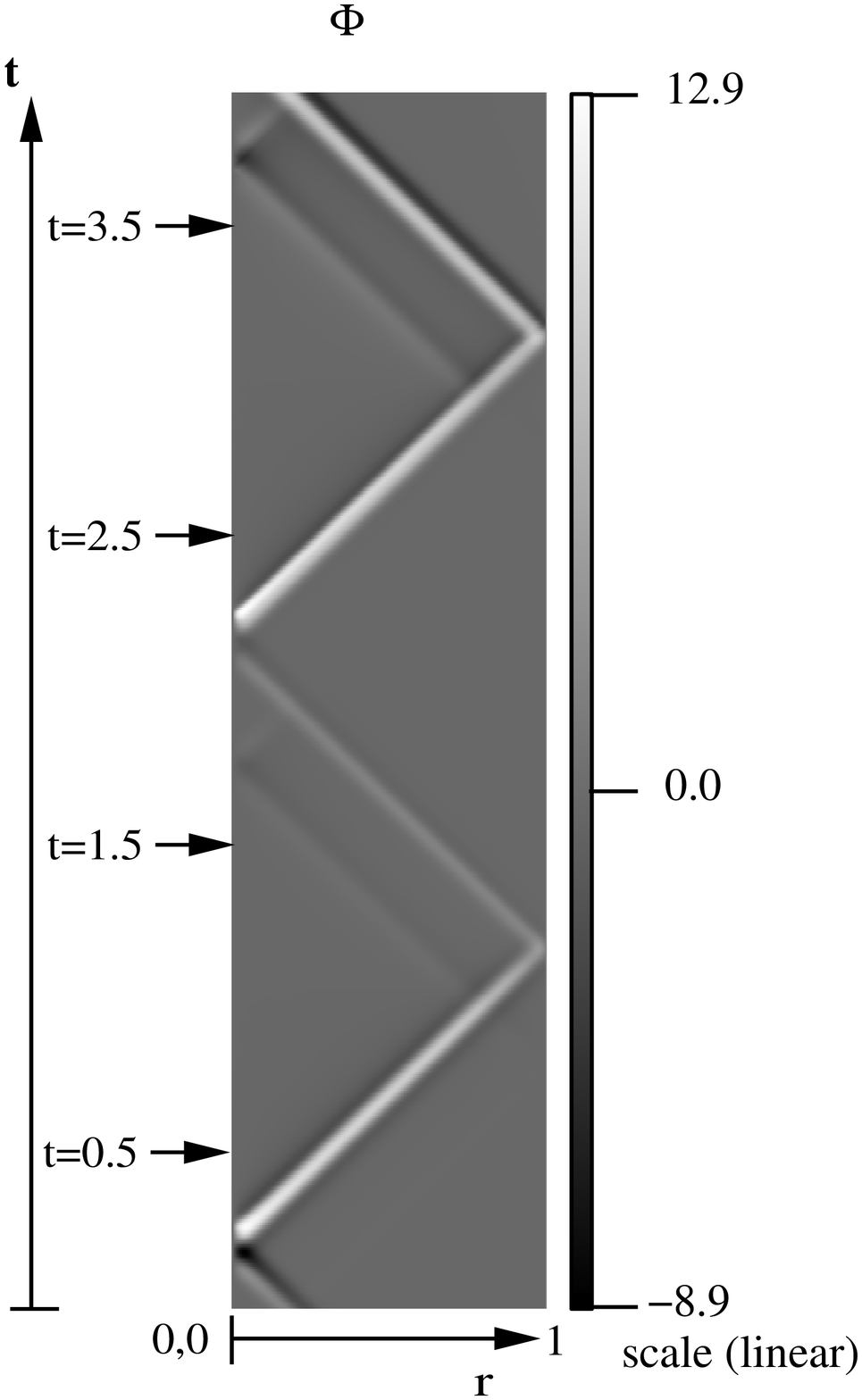}}
\caption{\label{g1_phi_1302_bw}
A plot of $\Phi(r,t)$, the spatial gradient of the scalar field, 
for sample gaussian initial data with 
$\amp=0.1302$. 
In this case, a black hole is {\em not} formed.
This plot clearly demonstrates
the nature of the Dirichlet boundary conditions on $\phi$ at $\Scri$ ($r=1$
in these coordinates). Even though a black hole does not form, 
back reaction {\em is} significant here---notice the 
non-linear interaction between ingoing
and outgoing components of the field:
when the ingoing and outgoing pulses cross, the ingoing component
is amplified, while at the same time the outgoing component is surpressed.
The effect is most apparent on this plot at around $t=3$ near
the outer boundary; and note that the initial outgoing component 
of the field is quite small and not visible in the picture.
}
\end{figure}

\subsection{Parameter space survey, varying $\amp$}

Figs. \ref{gmass} and \ref{cmass} show plots
of the asymptotic mass, $M(\amp)$, of the spacetime, as a function of 
the amplitude $\amp$, about the region $M=0$ of parameter space, for the 
guassian and harmonic families. The second curve on each plot shows the initial
mass estimate of a black hole (if one formed during the 2 LCTs of the gaussian
evolution, or 50 LCTs of the harmonic evolution) at the time an apparent horizon 
is first detected.  For these amplitudes, Figs. \ref{gr0t0} and \ref{cr0t0} show 
the time $t$ and coordinate position $r$ of apparent horizon formation.  
Qualitatively, the features of corresponding plots for the kink and
squared guassian (also evolved for 2 LCT's) are very similar to 
those for the gaussian, 
so for brevity we do not show them. 
To within the resolution of our simulation,
the final black hole mass always approaches the asymptotic mass---in other 
words, we do {\em not} detect any remnant scalar field (black hole `hair'). 
See Fig. \ref{g13am} for typical examples. 
Due to the ``reflecting'' boundary conditions at time-like $\Scri$, 
this is not too surprising, although 
one might have expected something like a low amplitude, long wave-length, 
periodic scalar remnant.
The scalar field also tends to zero
at late times along the event horizon, though in that region of the spacetime 
our results are not good enough to obtain useful decay exponents. 

\begin{figure}
\epsfxsize=17cm
\centerline{\epsffile{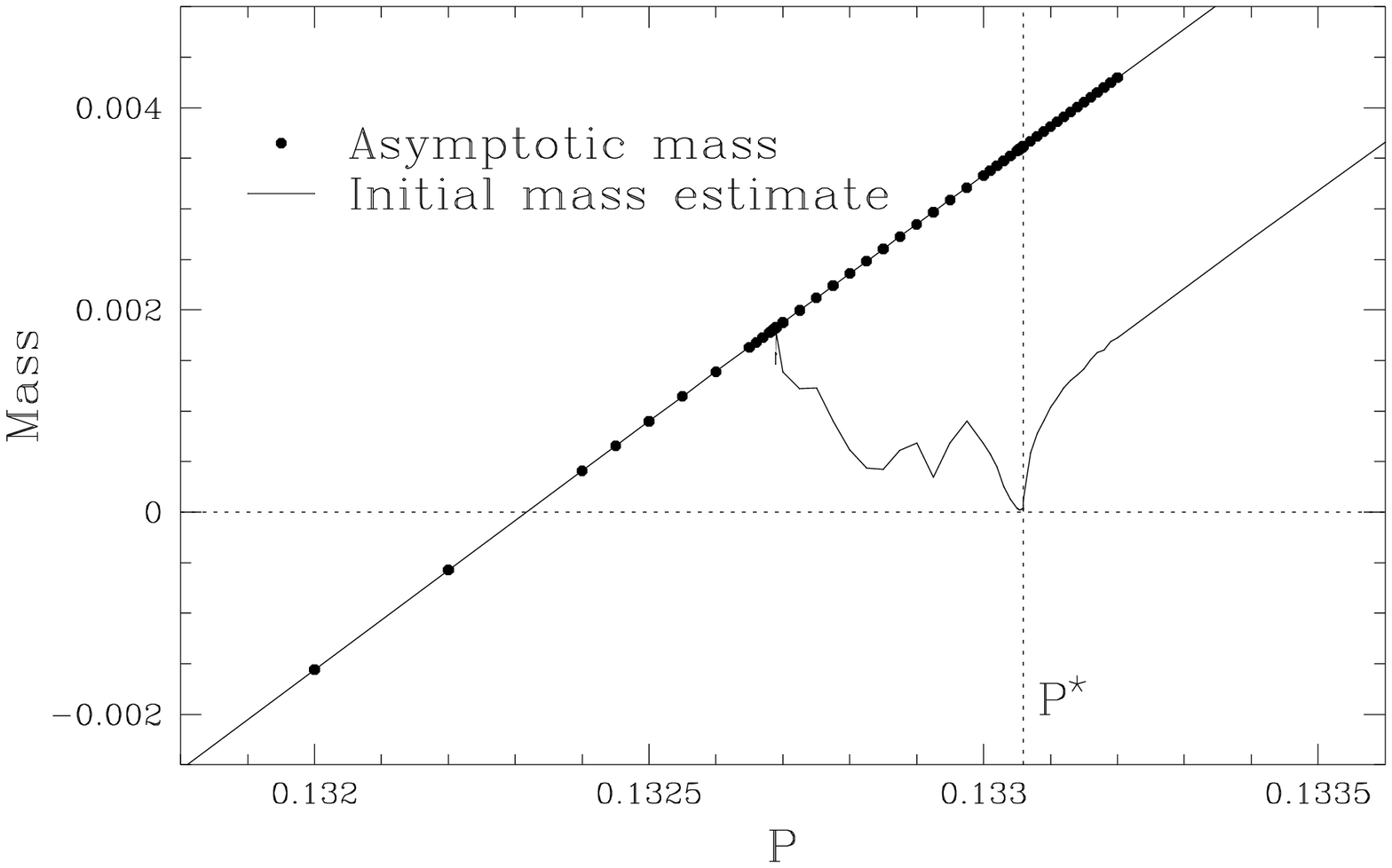}}
\caption{\label{gmass}
Asymptotic mass as a function of pulse amplitude for an
initially ingoing gaussian (\ref{gauss}) of width $0.05$, centered
at $r=0.2$ in a cosmology with $\ell=2/\pi$. For the amplitudes
that formed an apparent horizon within the simulation time of $t=2$,
the mass estimate at time of AH formation is also shown (it is
not clear in the figure but this curve does not touch
the asymptotic mass curve). 
The dashed vertical line, labeled by $\amp^\star$, 
is the critical amplitude---see sec. \ref{sec_crit}. }
\end{figure} 

\begin{figure}
\epsfxsize=17cm
\centerline{\epsffile{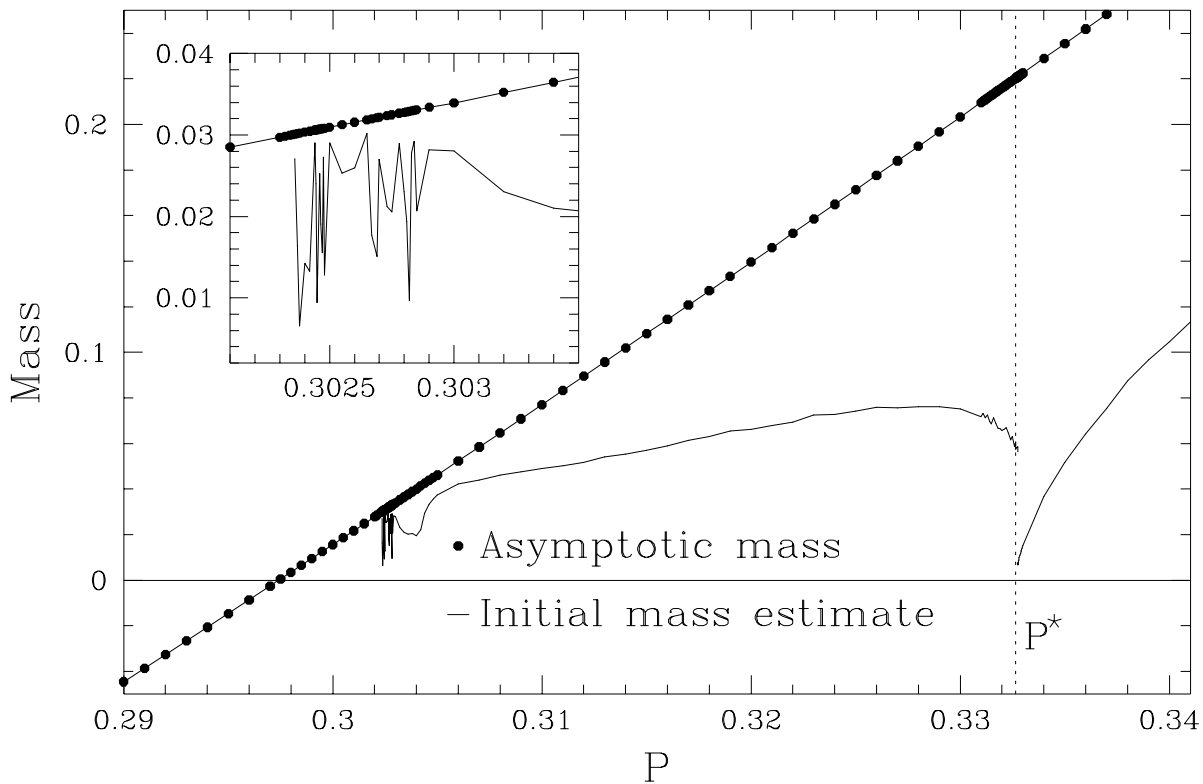}}
\caption{\label{cmass}
Asymptotic mass as a function of pulse amplitude for the time-symmetric
$n=1$ harmonic function (\ref{harm}).  For the amplitudes
that formed an apparent horizon within the simulation time of $t=50$,
the mass estimate at time of AH formation is also shown.
The dashed vertical line, labeled by $\amp^\star$, is the critical amplitude
as discussed in sec. \ref{sec_crit}. 
Notice the discontinuity of the initial mass estimate curve just to the right
of $\amp^\star$, and compare the gaussian case in
Fig. \ref{gmass}. The reason for the sudden jump, and difference
from the gaussian case, is that 
around $t=1$ for those amplitudes near $\amp^\star$ an apparent horizon is 
close to forming in two locations; to the left of the discontinuity it first 
forms at larger radii, to the right at smaller (see Fig. \ref{cr0t0}).
}
\end{figure} 

\begin{figure}
\epsfxsize=17cm
\centerline{\epsffile{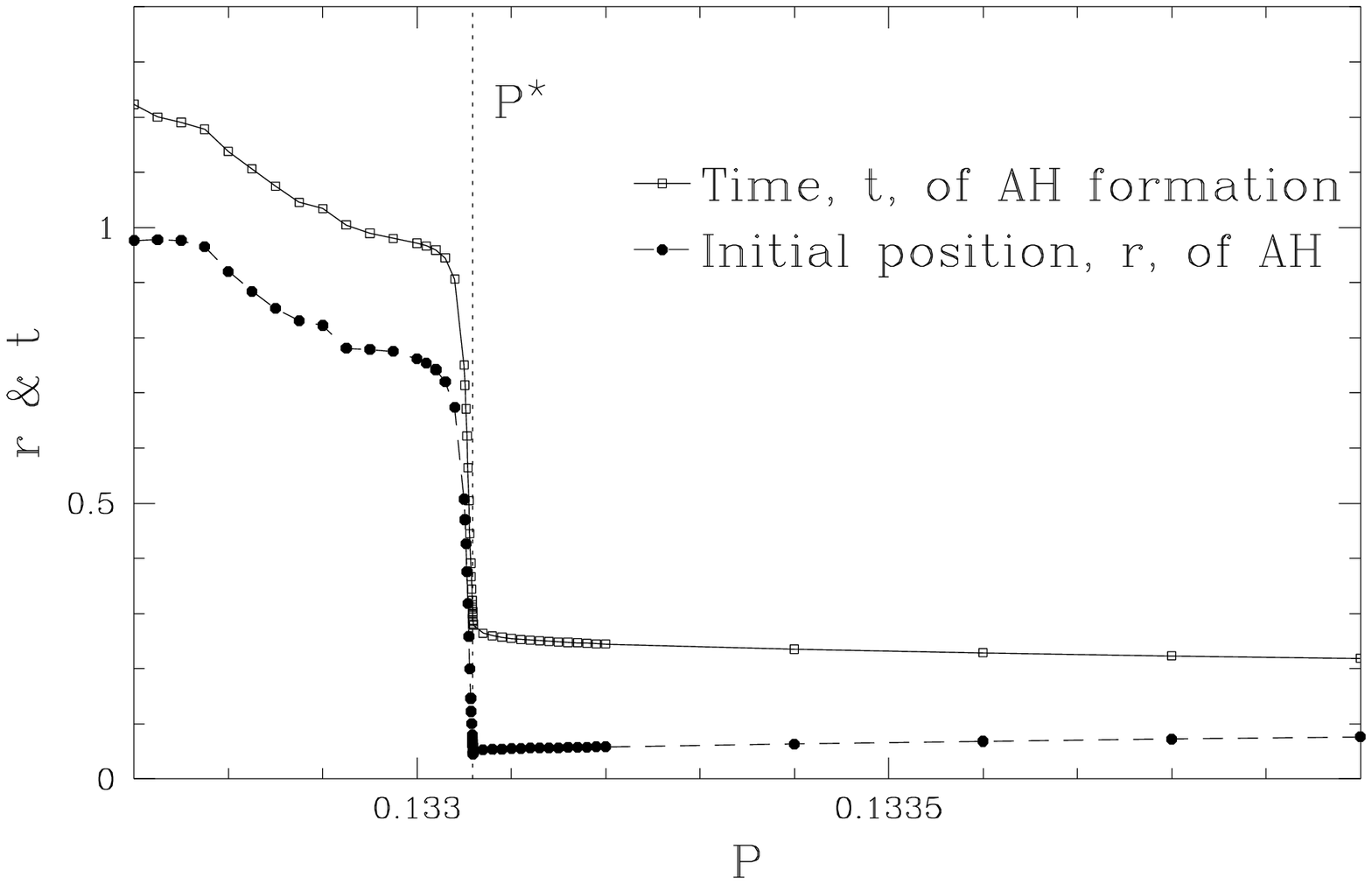}}
\caption{\label{gr0t0}
The initial coordinate position ($r$) and time ($t$) of AH formation
for the same set of amplitudes as in Fig. \ref{gmass} for the 
gaussian family (if an AH formed within $t=2$). }
\end{figure}      

\begin{figure}
\epsfxsize=17cm
\centerline{\epsffile{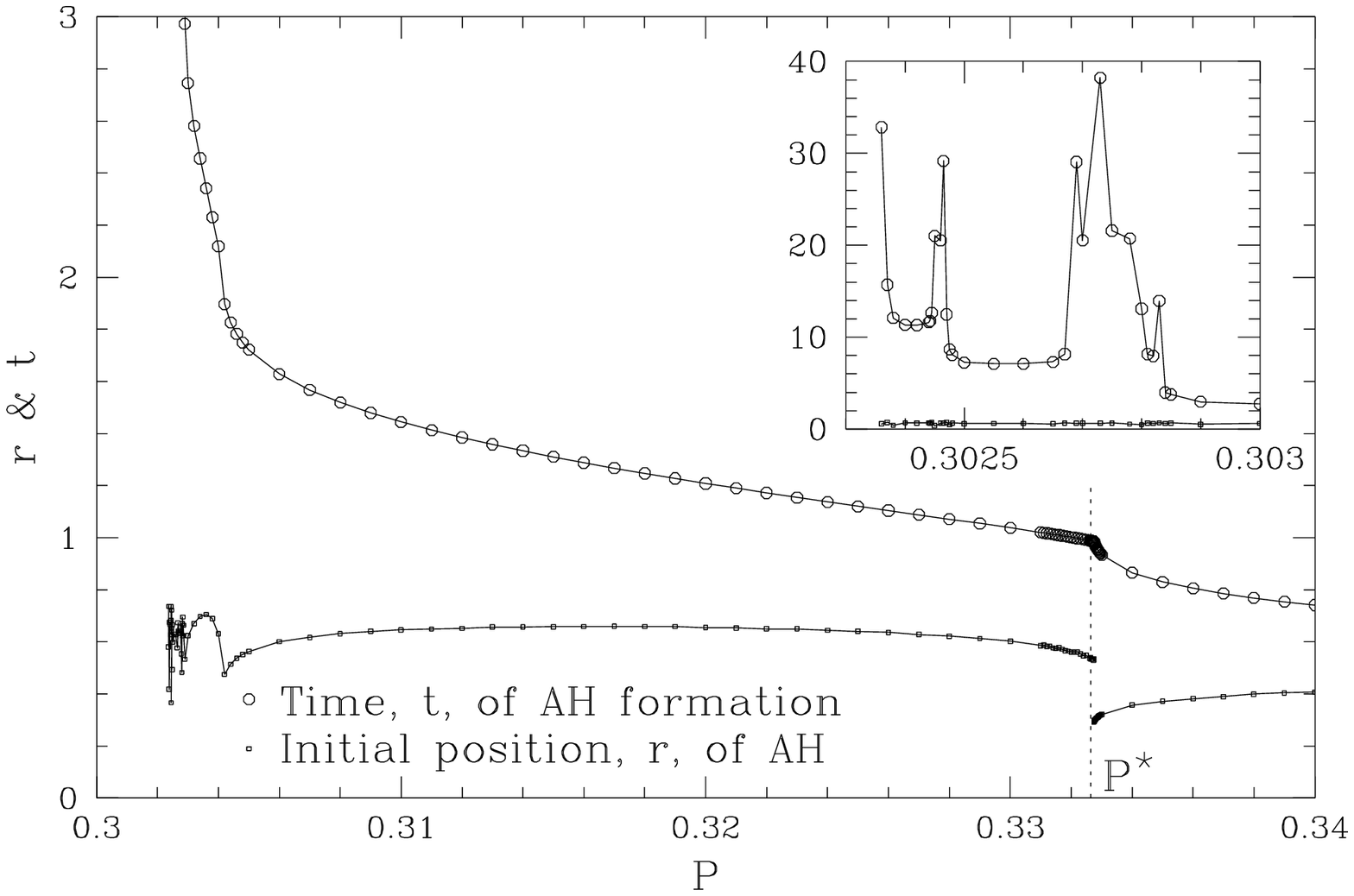}}
\caption{\label{cr0t0}
The initial coordinate position ($r$) and time ($t$) of AH formation
for the same set of amplitudes as in Fig. \ref{cmass} for the
harmonic familty (if an AH formed within $t=50$).
}
\end{figure}      

\begin{figure}
\epsfxsize=17cm
\centerline{\epsffile{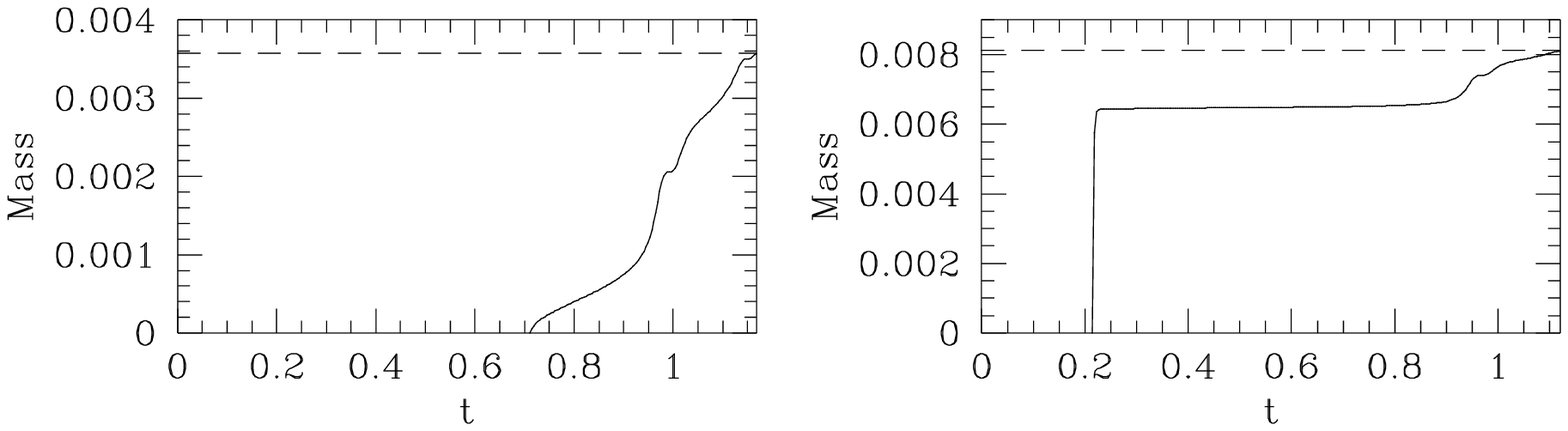}}
\caption{\label{g13am}
Black hole mass estimates as a function of time for 
gaussians with $\amp=0.133051$ (left) and $\amp=0.1340$ (right).
The horizontal dashed lines denote the asymptotic masses of the spacetimes.
For the less massive pulse on the left, the apparent horizon forms after the
initial implosion when the field is mostly outgoing, and the energy gradually
accretes onto the black hole.
The more massive pulse on the right forms a black hole on the initial
implosion, capturing almost all of the scalar field energy except
for a small piece that initially traveled outwards from $t=0$. This piece
eventually bounces off $\Scri$ then falls into the black hole. 
Note that the smaller amplitude is less than the critical value,
$\amp^\star$, as defined in sec. \ref{sec_crit}, while the larger
amplitude is super-critical. 
}
\end{figure}         

Fig. \ref{cr0t0} for the harmonic family shows almost chaotic dependence of 
the time of AH formation as a function of amplitude, as $M(\amp)$ decreases 
towards $M=0$. There is evidence that this behavior is also present for the 
other families of initial data, but we have not run those simulations 
at the necessary resolution to give {\em convincing} evidence. 
What appears to be happening is the following.
First of all, it is more ``difficult'' for a 
distribution of the scalar field corresponding to $M \gtrsim 0$ 
to form a black hole---the 
distribution needs to be compact and centrally condensed.
Thus, when we implode a relatively 
``space-filling'' distribution with $M$ small (and positive) a black hole
will not form on the first bounce through the origin. 
However, because of the boundary conditions at infinity,
the scalar field will reflect off $\Scri$, and, 
as the field has evolved through a strong
field (non-linear) regime in the interior, the distribution of energy will be
different on the subsequent implosion. 
Moreover, because of the strong gravitational field, the scalar field 
has a tendency to spend more time in the vicinity of the origin on average, 
preventing it from dispersing throughout the spacetime
. So, one may expect that
if the asymptotic mass $M$ is positive, a region of phase space will 
eventually be traversed during
evolution, where it is favorable for a black hole to form, no matter how 
near-zero is $M$.
However, due to the chaotic nature of the curve in Fig. \ref{cr0t0},
we cannot
extrapolate $t_0(M)$ to $t_0=\infty$ in order to directly 
test this conjecture.

\subsection{The critical regime}\label{sec_crit}

To search for critical behavior in the gravitational collapse of the four
families of initial data introduced in the previous section, we vary the
amplitude $\amp$ in each case to find the threshold of black hole formation. 
Ideally, we would simply 
seek the amplitude $\amp^\star$ where a black hole forms
for $\amp>\amp^\star$, while for $\amp<\amp^\star$ the 
scalar field bounces around 
forever without collapse. 
Unfortunately, such a search is not practical; as mentioned in the
previous section we do not have the computational resources to follow
compact initial data for numerous LCT's, and, even with the $\cos^2$ data,
we do not see any trends that would allow us to conclude that if a black hole
has not formed after, say, $n$ LCT's, then it probably will not form at all. 
Thus, what we do instead
is tune to the threshold of black hole formation on the initial implosion; i.e. 
we base our search on whether
or not a black hole forms {\em before} any initially out-going radiation 
reflects off $\Scri$ and then falls in, contributing to the collapse. 
This point of parameter space is labeled as $\amp^\star$ 
in Figs. \ref{gmass} and \ref{cmass}, and coincides with the place where
the initial mass estimate dips to near zero 
(though for the harmonic data---as 
mentioned in the caption of Fig. \ref{cmass}---for amplitudes
a little larger than $\amp^\star$ an apparent
horizon first forms further out, engulfing the one that is about to form at the 
smaller radius; see also Fig. \ref{cr0t0}).

Near this threshold, it turns out that
shortly after the initial implosion, 
the scalar field and geometry close
to the origin evolve
towards a universal, continuously self-similar (CSS) form.
We remind the reader that a function which is CSS depends only on a single 
scale-invariant variable $x$.  
Now, the coordinates $(r,t)$ in which we solve the equations
of motion
are {\em not} well-adapted to self-similarity. However, after some 
experimentation we found
that a natural scale-invariant independent variable in our system is
\begin{equation}\label{xdef}
x=\frac{\rb}{t_c},
\end{equation}
where $\rb=\ell\tan(r/\ell) e^B$ is proportional to the proper 
circumference of an $r=\constant$ ring, and $t_c$ is 
proper time as measured by the central ($r=0$) observer. By convention, $t_c$
is negative and increases to the accumulation point $t_c^\star \equiv 0$. 
To better visualize the CSS behavior, we also transform to logarithmic 
coordinates:
\begin{equation}\label{logdef}
Z \equiv \ln(\rb), \ \ \ T \equiv -\ln(t_c).
\end{equation}
A CSS function, $f(x)=f(e^{Z+T})$, then looks like a wave propagating to the left
with unit velocity as $T$ increases to $\infty$.

Figs. \ref{g1_phi_crit}--\ref{g1_r_crit} show scale-invariant functions 
$\phi_{,Z}(Z,T)$, 
$\phi_{,ZZ}(Z,T)$ (the second derivative
better demonstrates the ``wave nature'' of the critical solution),
the mass aspect $M(Z,T)$, and $(\rb^2 R)(Z,T)$,
for a gaussian evolution with $\amp=0.133059219$, which is `close' 
to the critical solution ($\ln(\amp-\amp^\star)=-17.5$; see 
sec. \ref{sec_gamma}).  In principle, the 
closer to criticality we tune the initial pulse, the longer the
scale-invariant behavior should persist in logarithmic space. 
In practice, of course, finite computational 
precision and grid resolution prohibits 
fine-tuning to arbitrary accuracy---the figures plotted here show data
which is about as close to criticality as we can get with 8192 grid points. 
In terms of the mass aspect in Fig. \ref{g1_mass_crit}, one can surmise 
that the critical solution is (locally) a kink-like 
transition from the AdS value $M=-1$ to a
zero mass state; though, interestingly enough, 
the value of the curvature scalar $R$ at the
origin diverges like $1/t_c^2$ as one approaches the 
accumulation point (we will discuss this in more detail below; 
also, bear in mind that in Fig. \ref{g1_r_crit} we are plotting
$\rb^2 R$, not $R$ itself).
This behavior of the mass aspect {\em suggests} that the 
the transition at the critical point is Type II---in other words, 
there is no lower, positive bound
on the mass of black holes that can be formed by the scalar field. 

\begin{figure}
\epsfxsize=17cm
\centerline{\epsffile{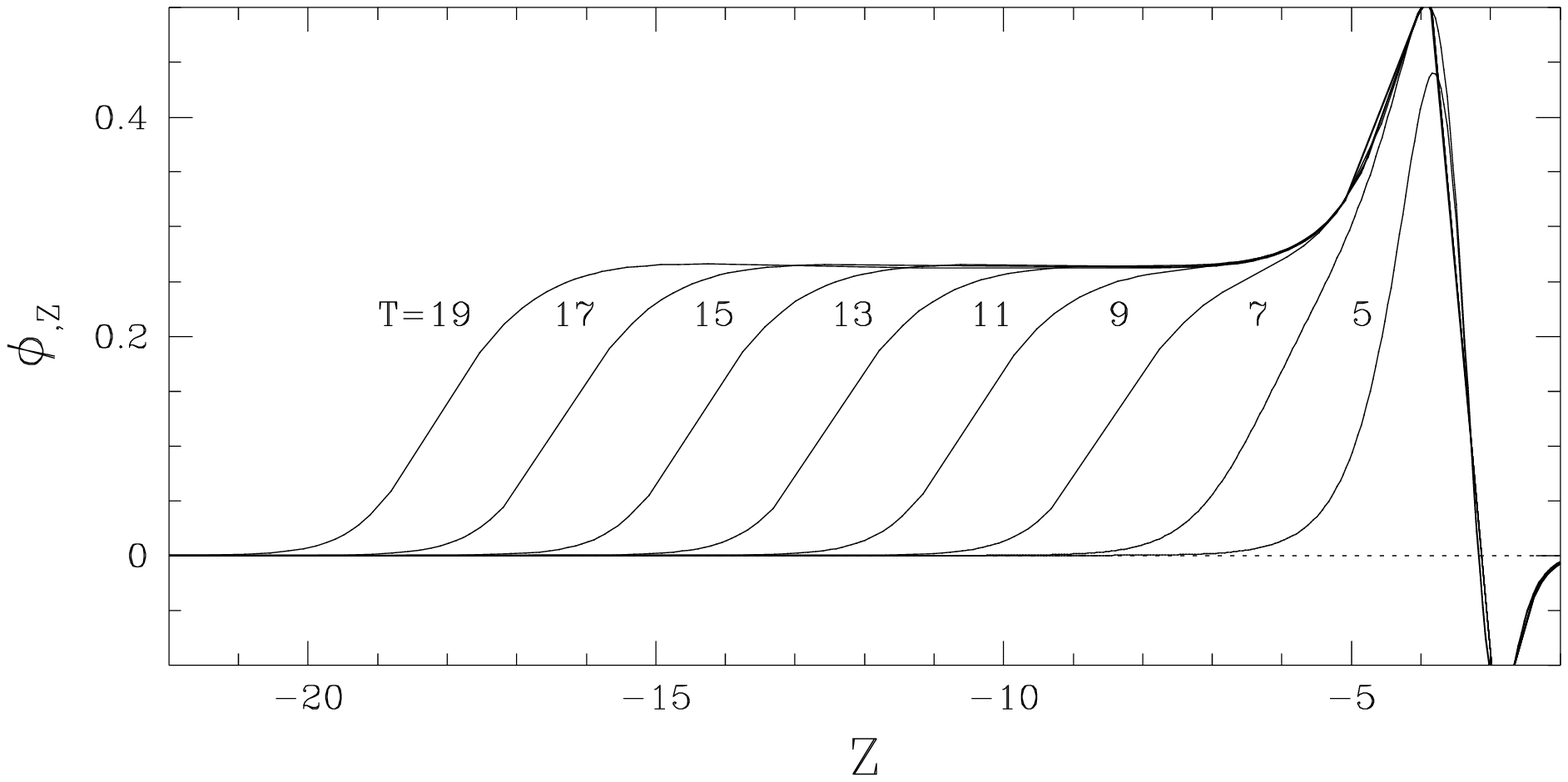}}
\caption{\label{g1_phi_crit}
$\phi_{,Z}(Z,T)$ (see (\ref{logdef}) for the definition of 
$Z$ and $T$ coordinates),
for gaussian initial (\ref{gauss}) data with $\amp=0.133059219$, $\width=0.05$,
$n=1$ and $\ro=0.2$ in an $\ell=2/\pi$ cosmology. This function of $\phi$
is scale-invariant in the critical regime, which unfolds 
roughly between $T\approx 8$ and $T\approx 19$ (though, 
interestingly, the scale-invariance seems to persist for longer
in the scalar field than the geometric 
quantities---see Figs. \ref{g1_mass_crit} and \ref{g1_r_crit}). 
}
\end{figure}

\begin{figure}
\epsfxsize=17cm
\centerline{\epsffile{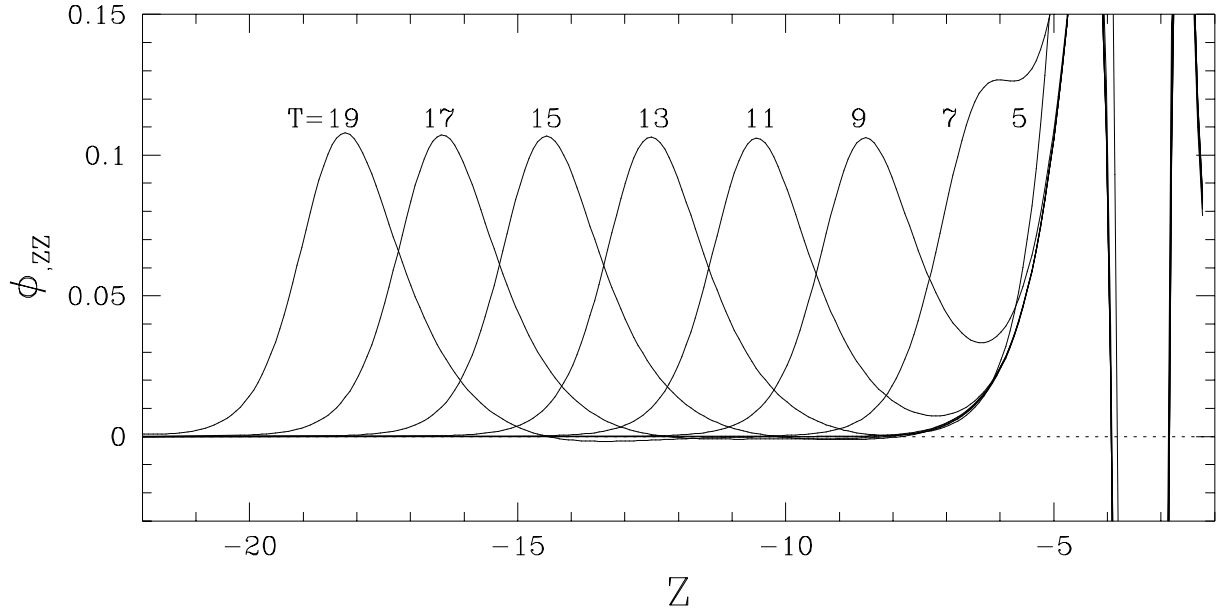}}
\caption{\label{g1_dphi_crit}
$\phi_{,ZZ}(Z,T)$, i.e.
the derivative of the function plotted in Fig. \ref{g1_phi_crit}.
}
\end{figure}

\begin{figure}
\epsfxsize=17cm
\centerline{\epsffile{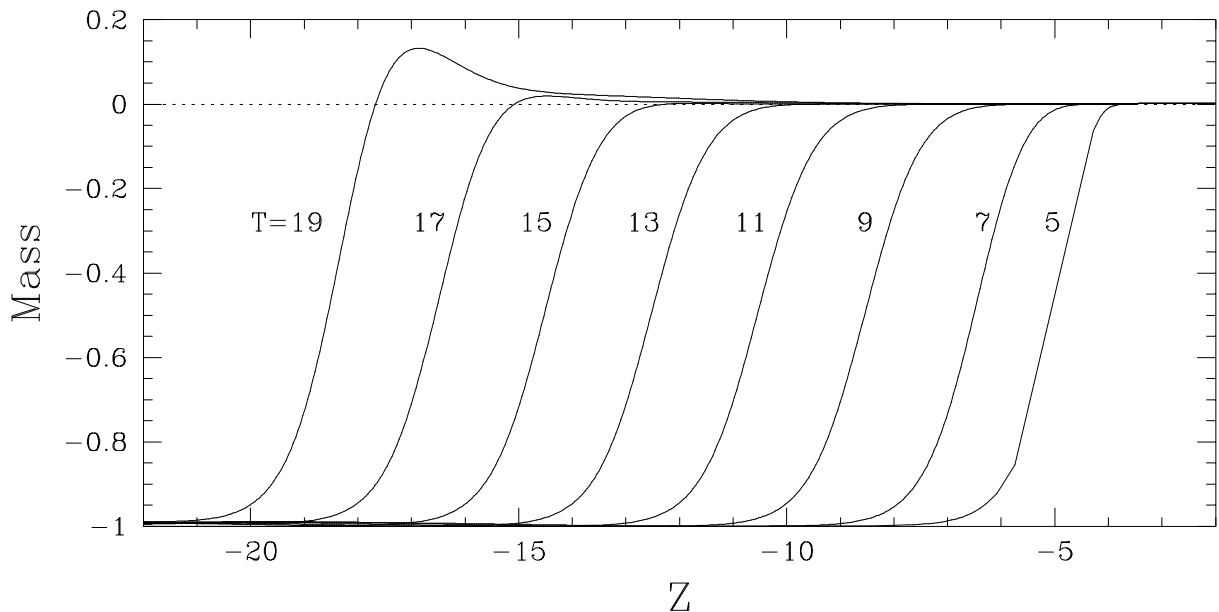}}
\caption{\label{g1_mass_crit}
The mass aspect, $M(Z,T)$, for the same solution shown in Fig. \ref{g1_phi_crit}. 
That $M$ becomes slightly non-monotonic at late times is probably due to 
numerical error---this is a super-critical evolution, and the metric
variables are already growing rapidly around $T=19$.
}
\end{figure}

\begin{figure}
\epsfxsize=17cm
\centerline{\epsffile{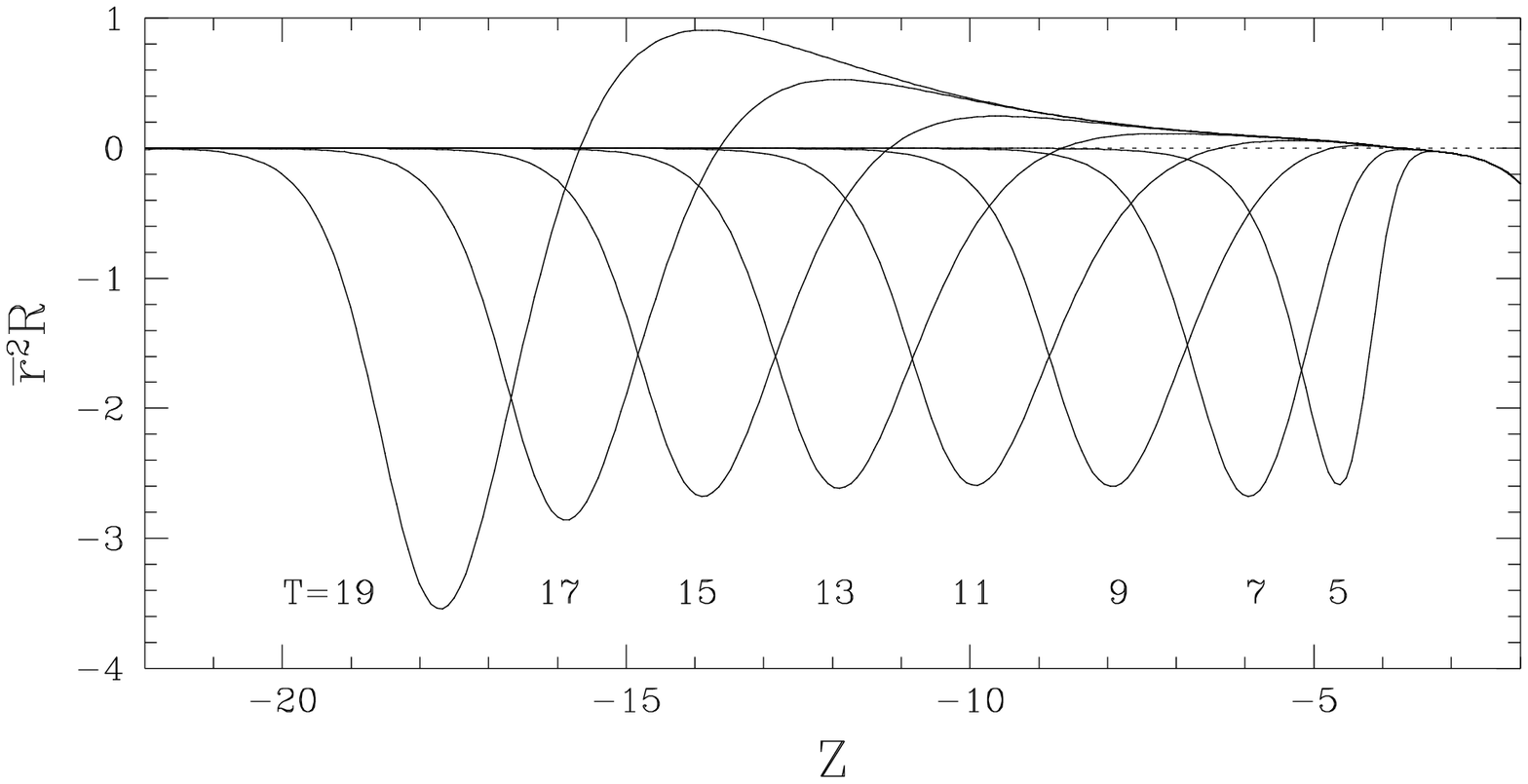}}
\caption{\label{g1_r_crit}
The Ricci scalar multiplied by $\rb^2$ (a scale-invariant 
combination in the critical regime)
for the same solution shown in Fig. \ref{g1_phi_crit}.
}
\end{figure}

Fig. \ref{comp} demonstrates the universality of the solution in 
the critical regime.
Here we plot $\phi_{,ZZ}$
(as in Fig. \ref{g1_dphi_crit} for the gaussian)
at the same time $T$ for each family in a near-critical evolution.
The harmonic function appears to have a slightly larger amplitude, but, as
we shall now argue, this is apparently just a slicing effect. 
As mentioned in sec. \ref{sec_ic}, because of the gauge that we
use, and since we choose to solve for $A(r,0)$ and $A_{,t}(r,0)$ using
the constraint equations (\ref{hamil}) and (\ref{momen}), 
the only slicing freedom we have remaining is in the
initial conditions for $B(r,0)$ and $B_{,t}(r,0)$.  Once $B(r,0)$ 
and $B_{,t}(r,0)$ are specified, we have no control over the
manner in which the slice evolves. For the three compact, ingoing families, 
the critical behavior develops at times ranging from $t=0.25$ to $t=0.30$, 
and because of the similar initial spatial distribution of the energy densities, 
the evolution has proceeded along very similar slices. On the other hand, 
the harmonic data approaches the critical solution at about $t=1.25$,
at which point the slice has evolved quite differently from the other three
families near their respective critical times (we note, however, that by plotting
as a function of $T$ we {\em do} ``match'' the slices at the origin). 
To demonstrate that the slices evolve differently, 
we plot in Fig. \ref{gamma} the normalized inner product between 
$\partial/\partial  t_c $ (in an $\rb,t_c$ coordinate basis) and $\nabla t_c$ 
for the 4 families, at the same 
time used in Fig. \ref{comp}. This inner product is the Lorentz gamma factor, $W$
(assuming the vectors are time-like), between $\rb=\constant$ observers, and 
those moving normal to the hypersurface $t_c=\constant$:
\begin{equation} \label{W}
W=\frac{|(\partial/\partial t_{c})^\alpha \nabla_{\alpha} t_c|}
{|\partial/\partial t_{c}| |\nabla t_c|}.
\end{equation}
This quantity 
will be the same along
identical slices of a spacetime 
(since such slices will have the same normal vectors); 
thus the harmonic solution slice is
clearly different as one moves away from the origin. Another interesting 
feature of this plot for the harmonic data is that it shows
gravitational collapse occurring a short distance away from the
unfolding critical behavior, since, to the right of the peak, the vector 
$\partial/\partial t_c$ has become space-like (equivalently the 
surface $\rb=\constant$ has become space-like---see the
discussion on the singularity structure in sec. \ref{sing}, and
in particular Fig. \ref{bh13305_prop_bw}). At this point in parameter space
for the harmonic function there 
is a lot more mass in the spacetime than that involved in the critical 
evolution, and this is causing an apparent horizon to form at a 
larger radius (see Figs. \ref{cmass} and \ref{cr0t0}). 
Presumably, for smaller amplitudes one could tune to a threshold solution
after several light-crossing times, and perhaps then one would more cleanly
uncover the critical solution. \par
To give more evidence that all the solutions are indeed approaching a 
universal one in the critical regime, we need to compare them
on a common spacetime slice. In Fig. \ref{comp_vrb} we show the same
function of the scalar field as in Fig. \ref{comp} transformed to a
Christodoulou type coordinate system ($\rb,v$), where a $v=\constant$
curve is an ingoing null geodesic \cite{chr}. We normalized $v$ 
so that $dv=dt_c$ at the
origin; i.e. $v$ also measures central proper time. Thus comparing
solutions on the same $v=\constant$ surface removes any slicing ambiguity 
\footnote{we are grateful to David Garfinkle for suggesting 
this procedure to us}.  As can be seen from the figure, the transformed 
solutions are all quite similar, though 
we lose some accuracy in the transformation (which is why we have
elected not to use these coordinates in all of the plots in Figs. 
\ref{g1_phi_crit} - \ref{comp}).

\begin{figure}
\epsfxsize=17cm
\centerline{\epsffile{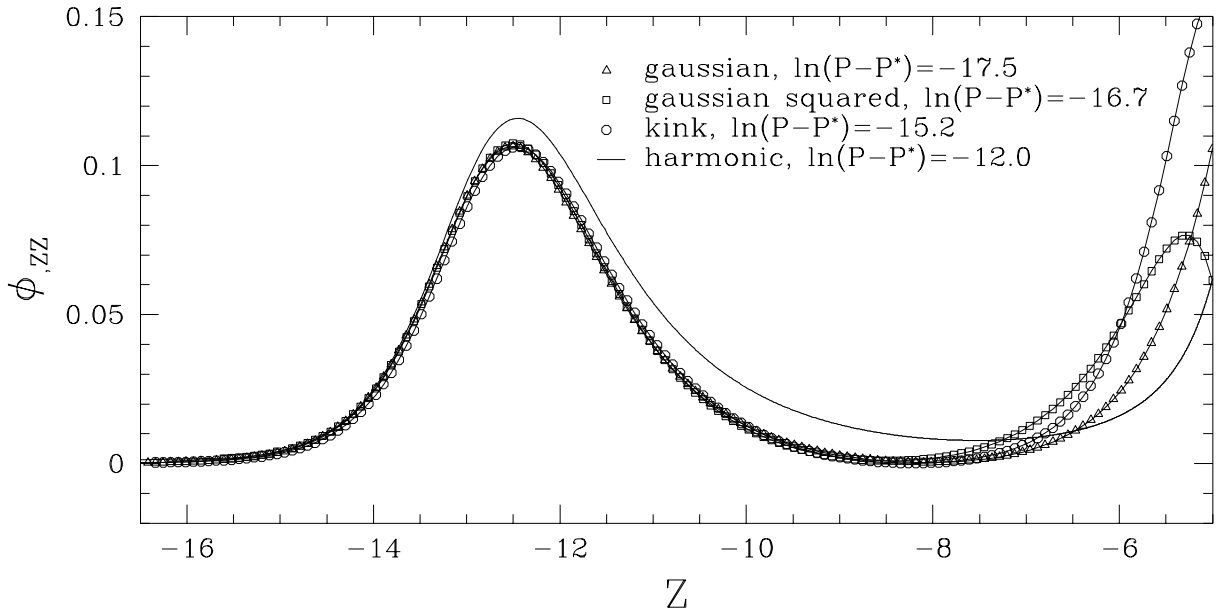}}
\caption{\label{comp}
A composite of the scale-invariant 
function $\phi_{,ZZ}(Z,T)$ for the near critical solutions
of the four families of initial data considered, demonstrating
universality of the solution in the critical limit. The data is plotted
at $T=13$ (compare Fig. \ref{g1_dphi_crit}). 
The harmonic profile appears somewhat different than the others 
due to a slicing effect, as
explained in the text (also see Figs. \ref{gamma} and \ref{comp_vrb}).
See Fig. \ref{scaling} for the values of $\amp^\star$ for each family.
}
\end{figure}

\begin{figure}
\epsfxsize=17cm
\centerline{\epsffile{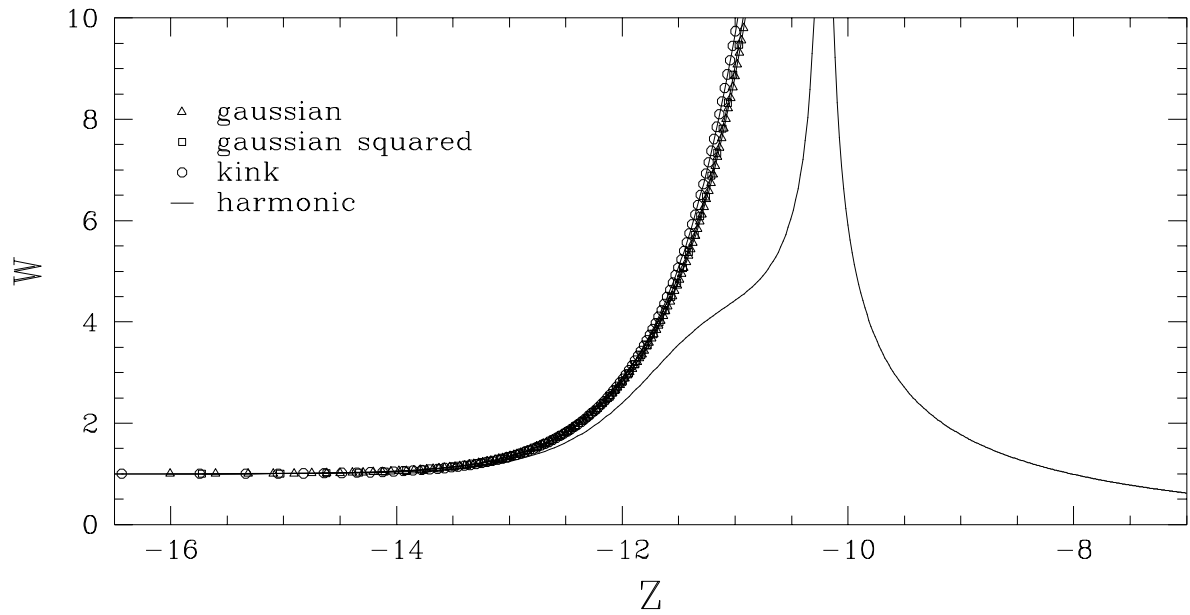}}
\caption{\label{gamma}
The Lorentz gamma factor W (\ref{W})
between $\rb=\constant$ observer world-lines
and those travelling normal to the hypersurface $t_c=\constant$, at $T=13$
for the 4 near-critical solutions as in Fig. \ref{comp}. The difference
between the three initially ingoing families and the harmonic one indicates that
we are looking at differing slices of spacetime as we move away from 
$r=0$. In fact, the discontinuous peak in $\gamma$ for the harmonic solution
at around $Z=-10.2$ shows that the $\rb=\constant$ 
surface has become space-like to the
right of this point, indicating a region undergoing gravitational collapse.  
}
\end{figure}

\begin{figure}
\epsfxsize=10cm
\centerline{\epsffile{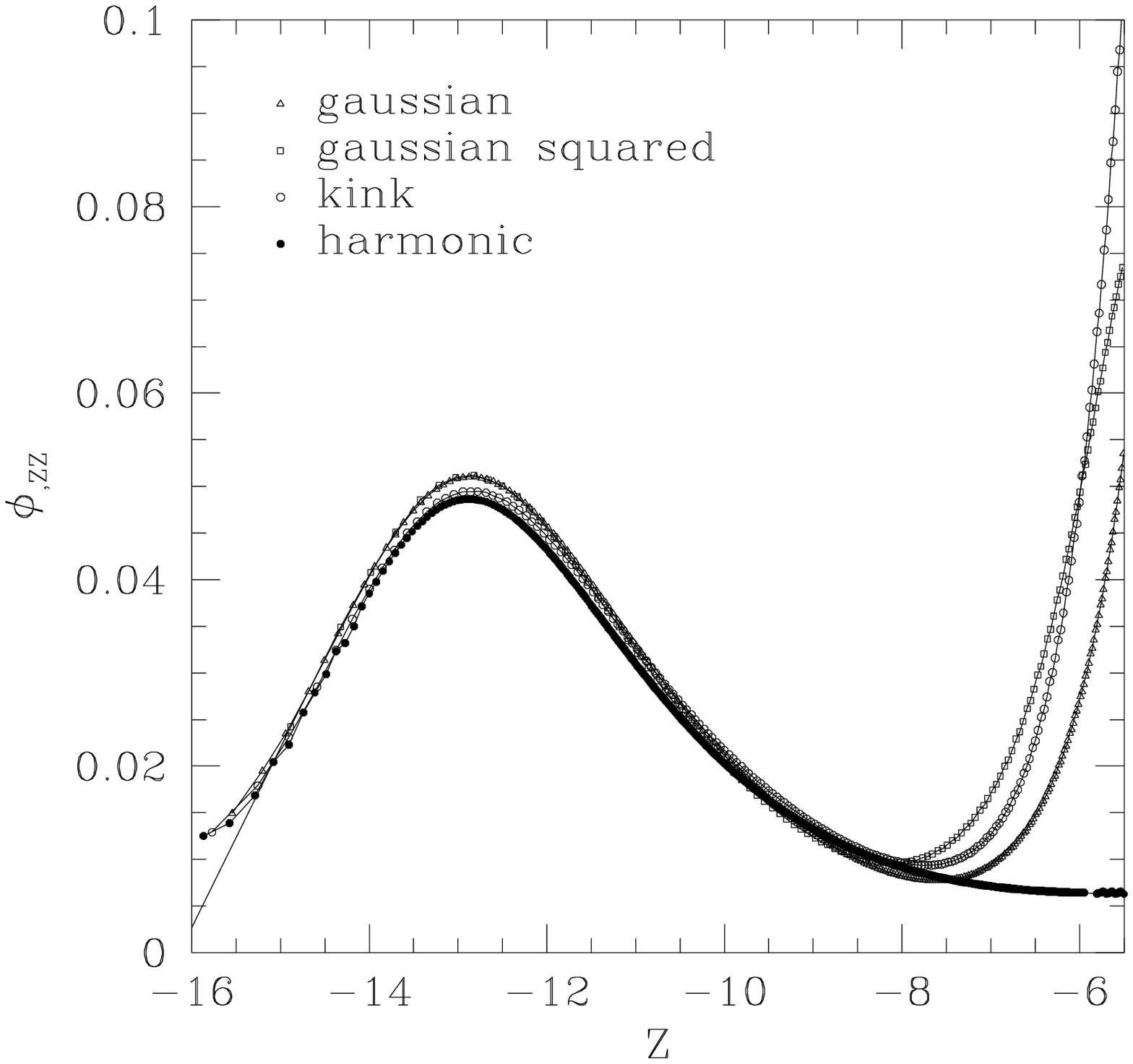}}
\caption{\label{comp_vrb}
The function $\phi_{,ZZ}(Z,v)$ for the same solutions shown 
in Fig. \ref{comp}. $v=\constant$ is an ingoing null geodesic, 
chosen here to intersect
the origin at $T=-\ln t_c=13$ in all cases, and along this hypersurface 
we plot as a function of $Z=\ln \rb$. This coordinate system completely
fixes the spacetime slice along which we are comparing solutions 
(at the expense of some loss of accuracy
during the transformation), and gives additional evidence that there {\em is} a
universal critical solution.
}
\end{figure}

\subsubsection{The scaling exponent $\gamma$}\label{sec_gamma}

Another characteristic feature of Type II critical behavior in 
gravitational collapse
is the universal scaling exponent $\gamma$ in the relation 
$M=K(p-p^\star)^{2\gamma}$.
To measure this relationship in the current context, one needs to 
wait for the system to settle
down to a steady-state to ensure that the apparent horizon is
coincident with the event horizon, and hence that the mass 
estimate (\ref{m_approx})
gives the correct mass. In AdS, the boundary conditions at 
$\Scri$ prevent us from performing this measurement---initially 
outgoing radiation that 
did not contribute to the near-critical black hole formation will eventually
reflect off $\Scri$ and pollute our measurement. However, as discussed by 
Garfinkle and Duncan \cite{garfinkle}, in the near-critical regime 
(above or below $p^\star$) any quantity 
with dimension $L^q$, where $L$ is a length scale, 
should exhibit a scaling relation with an 
exponent of $q \gamma$. Thus, following those authors, we find the maximum
value attained by the Ricci scalar $R$ at $r=0$ in sub-critical evolution for
$t_c<0$. Plots of 
$\max_{t<t_c} \ln |R(0,t)|$ vs. $\ln(\amp^\star-\amp)$ for the four
families studied is shown in Fig. \ref{scaling}. Since $R \propto L^{-2}$, 
these figures show that the scaling exponent $\gamma$ 
of the 2+1D AdS Klein-Gordon system is about $1.2 \pm 0.05$.

Notice that the mass
aspect $M$ as
defined in (\ref{Mdef}) is dimensionless (which is consistent with the
scale-invariance of $M$ as plotted in Fig. \ref{g1_mass_crit}). On the other
hand, when we keep $\ell$ fixed and vary $\amp$, the resulting black hole mass 
(being proportional to $r_{ah}^2$) has a length scale of $2$, so one would
expect the mass-parameter scaling relationship for BTZ black holes to go like
$M=K(\amp-\amp^\star)^{2\gamma}$, where $\gamma$ is the same value $1.15-1.25$
found above for the scaling of $R$. The initial-mass estimate curves
as shown in Figs. \ref{gmass} and \ref{cmass} {\em do} 
roughly exhibit this scaling behavior for $\amp>\amp^\star$.

\begin{figure}
\epsfxsize=17cm
\centerline{\epsffile{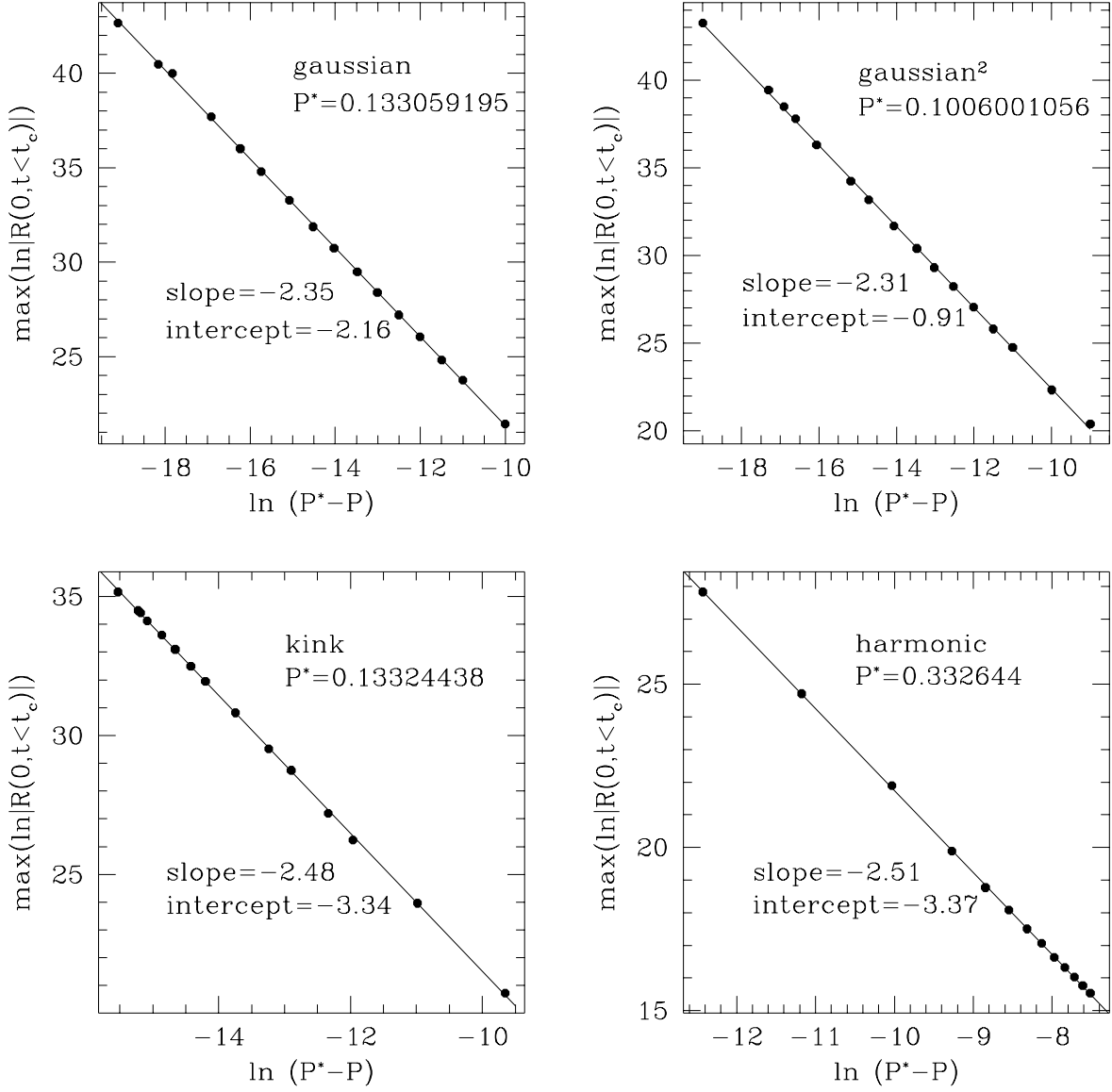}}
\caption{\label{scaling}
The maximum of $\ln|R(0,t<t_c)|$ as a function of $\ln(\amp^\star-\amp)$
for sub-critical ($\amp<\amp^\star$)
evolutions of the 4 families of initial data considered. These plots indicate
that the maximum of $R(0,t<t_c)$ attained during evolution is proportional to 
$(\amp^\star-\amp)^{-2\gamma}$, with $\gamma \approx
1.15-1.25$.
}
\end{figure}

\subsubsection{Critical behavior in the presence of a point
particle}\label{sec_pp}
Here we briefly show how the presence of a point particle (angle deficit)
alters the critical solution. The particle contributes to the
mass of the spacetime (\ref{m_pp}), so the more massive the 
particle (up to the maximum $\mpp=1$ in our units) the less scalar field energy
is needed to form a black hole, and consequently we have smaller amplitudes,
$\amp^\star$, at threshold. Interestingly, we find the {\em same} critical
solution in all cases (see Fig. \ref{comp_a00} for 3 examples), the only
noticeable differences being a systematic phase shift in $T$ related to the
mass of the particle. The kink-like transition in the mass aspect has the
same shape as well, but it ranges from the particle mass at $r=0$ to $M=0$. 
To within the resolution of our simulations
(which was at 2048 grid-points in this case) the critical exponent
is also the same, namely within the range $\gamma=1.15$ to $1.25$.

\begin{figure}
\epsfxsize=17cm
\centerline{\epsffile{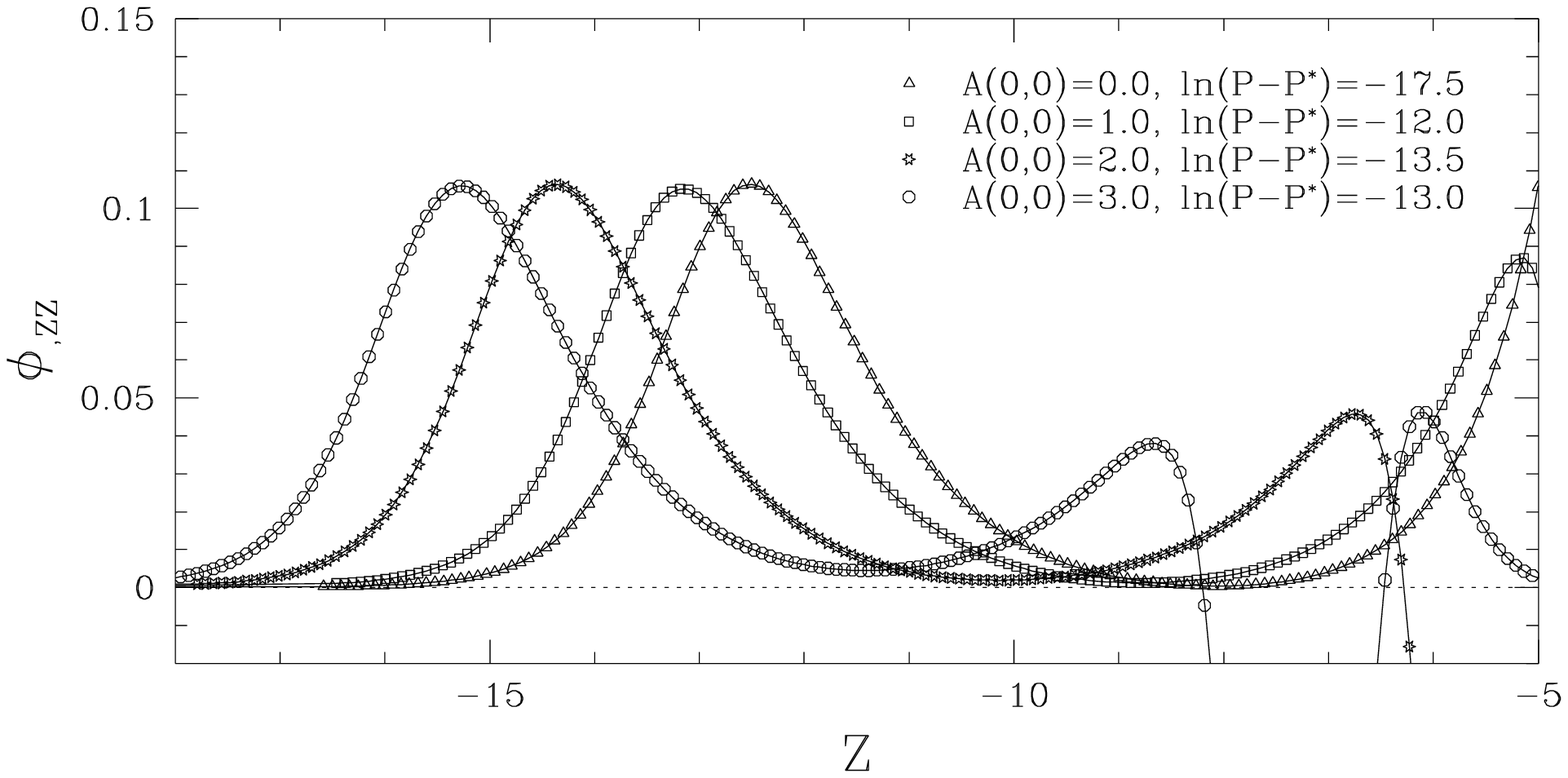}}
\caption{\label{comp_a00}
A composite of the scale-invariant function $\phi_{,ZZ}(Z,T)$ at 
$T=13$ for the near critical solutions
of the gaussian family ($\width=0.05, \ro=0.2$) with 4 different initial
values for $A(0,0)$---0, 1, 2 and 3, corresponding to 
the presence of point particles at the origin with masses 0, 0.864665, 
0.981684 and 0.997521 respectively (\ref{m_pp}). It is striking that these
solutions only differ by a phase in $T$ related to the particle masses; they
have evolved to the same amplitude after starting with quite different
initial amplitudes (namely, $P\approx 0.13, 0.078, 0.034, 0.013$ from
$A=0$ to $3$).
}
\end{figure}

\subsubsection{The critical solution from a CSS ansatz?}\label{sec_css}

Given that we have self-similar behavior in the critical regime, it would
be useful to find the exact solution assuming a CSS ansatz. Traditionally
this is done by assuming the existence of a homothetic Killing vector. $\xi$ 
(see \cite{cp_review}) 
\begin{equation}\label{homothetic}
\mathcal{L}_\xi g_{ab}=2 g_{ab}.
\end{equation}
This implies that in coordinates adapted to the homotheticity, so that
$\xi=\partial/\partial \tau$, each component
of $g_{ab}$ has the form $e^{2\tau} f$, 
for some function $f$ independent of $\tau$.
Furthermore, $\mathcal{L}_\xi R_{ab}=0$, so that $R_{ab}$ and hence the
Einstein tensor $G_{ab}$ are independent of $\tau$. This ansatz is not 
consistent with the field equations (\ref{ekg}) in the presence of the
cosmological constant if we assume that the
scalar field is self-similar (see \cite{brady}), as we observe in
the collapse simulations. Essentially, the
scalar field stress-energy tensor (\ref{kgset}) would need to decouple into a 
piece that exactly cancels the cosmological constant term plus a
scale-invariant term, but we do not think that this is possible for a
minimally-coupled scalar field.

It may be that in the 2+1D AdS system a different symmetry, such as a conformal
Killing vector, would be needed to generate the critical solution. Or perhaps 
the critical solution is only approximately homothetic over a limited region of 
the spacetime. Nevertheless, we have not yet found a symmetry-reduced system
that reproduces the observed critical behavior.\footnote{Note 
added in preparation:
David Garfinkle has very recently found a CSS solution in the limit 
where the cosmological
constant vanishes that appears to quite accurately describe the 
critical solution that we have found \cite{garfinkle2}. His result is
quite intriguing---the cosmological constant is {\em essential} for
black holes to form, yet apparently it plays very little role in the
solution at the {\em threshold} of formation! }

\subsection{Singularity structure}\label{sing}

In all of the solutions that we have studied so far we find that after an 
apparent horizon forms what appears to be a spacelike 
curvature singularity develops within the horizon. Specifically,
the surface of excision along which the metric variables $A$ and $B$
and, consequently, the curvature invariants
begin to diverge, is spacelike. By itself, demonstrating a spacelike surface
of arbitrarily large curvature is not sufficient to prove that the singularity
is spacelike---a counter-example would be the mass-inflation null singularity
\cite{null_sing} \footnote{We are grateful to Lior Burko for pointing this
out to us}. However, if we extrapolate to the surface of infinite curvature,
based upon the growth of the Ricci scalar prior to excision, we still
find a spacelike surface (in fact, $R$ grows so rapidly prior to excision
---roughly like $1/t^4$ along an $r=constant$ surface if we translate $t$ to
zero at the singularity---that the surface of infinite curvature 
essentially coincides with the excision surface at 
the resolution of Fig \ref{RS133051_bw} below). In addition, $B(t,r) \to
-\infty$ along this surface, indicating that the proper circumference measure
$\ell \tan(r/\ell) e^{B}$ goes to zero there (see Fig. \ref{bh13305_prop_bw}
below). Thus, 
as with vacuum BTZ black holes, this singularity is crushing\footnote{or {\em
deformationally strong}, see \cite{sing}. It is straight-forward (though
tedious) to see that $r=0$ in the non-rotating $BTZ$ black hole is a strong
singularity as defined by Tipler (though it is not a curvature singularity!). 
We have not repeated the formal calculations in terms of Jacobi fields in our 
collapse simulations, but because of the central, space-like nature of the
singularity back-reaction is not likely to weakening it. 
Note added in revision: shortly after this paper was first published, 
Lior Burko studied the
structure of the singularity in 2+1D AdS spacetime using a 
`qausi-homogenious' approximation, and did find 
the singularity to be strong and spacelike \cite{ads_sing}.}:
any extended object reaching the singularity is forced to zero proper
circumference, regardless of any angular momentum or internal pressures
that the object might have. 

Figs. \ref{PHI133051_bw} and \ref{RS133051_bw} are spacetime 
plots 
(essentially Penrose diagrams)
of $\Phi(r,t)$ and the Ricci scalar $R(r,t)$, respectively, 
for a gaussian initial pulse 
with $\amp=0.133051$. 
On the pictures we have superimposed the region 
of trapped surfaces and the inferred event horizon of the space
time, found by tracing a null ray backwards in time from the
place where the AH meets $\Scri$ on the coordinate grid. 
Fig. \ref{bh13305_prop_bw} show contours of proper circumference for the same 
solution.  The point $\amp=0.133051$ in parameter space is slightly 
sub-critical (as 
we have defined criticality, see Sec.~\ref{sec_crit})---a black hole forms
because the bit of outgoing energy present at $t=0$ bounces off $\Scri$ and
falls back onto the nearly collapsed scalar field, pushing it over the limit. 
This gives us a very clear view of the interior structure; for a more massive
pulse the singularity forms shortly after the initial implosion, resulting 
in a thin sliver of an interior in $(r,t)$ coordinates.

From Fig. \ref{RS133051_bw} one can see a
striking peak that forms in $R$ after the scalar field has bounced through
the origin and is travelling outwards.
In this particular case $R$ has a
value of order $-10^{10}$ in the interior, it then grows to order $+10^8$ 
over a very short distance before decreasing to the AdS value of $-6/\ell^2 
\approx -15$.  This near-discontinuous behavior in $R$ is characteristic of
sub-critical evolutions, and becomes more extreme as one nears the critical
solution.

As one approaches the excised space-like surface in Fig. \ref{RS133051_bw}, 
$R$ starts to grow very rapidly, 
reaching values up to $|10^{28}|$ before excision (this may not be 
clear on the figure---we chose the gray scale to highlight the 
near-discontinuous behavior in $R$). $R$ actually
oscillates between large positive and negative values along this surface,
but our calculations are not sufficiently accurate to conclude that the
oscillation is genuine. In particular,
$R$ is extremely sensitive to the difference $\Pi^2-\Phi^2$ (see (\ref{rs})),
and $\Pi^2$ is usually around the same order of magnitude as $\Phi^2$ there.
We also note that the maximum value attained by $R$ along the excised surface 
becomes smaller towards $\Scri$. This is to be expected, since in the 2+1D 
system, {\em some} scalar
field is necessary to produce a value of $R$ differing from the AdS value 
(again, see (\ref{rs})), and as we move towards $\Scri$ along the excised 
surface there is progressively less scalar field energy remaining.

\begin{figure}
\epsfxsize=16cm
\centerline{\epsffile{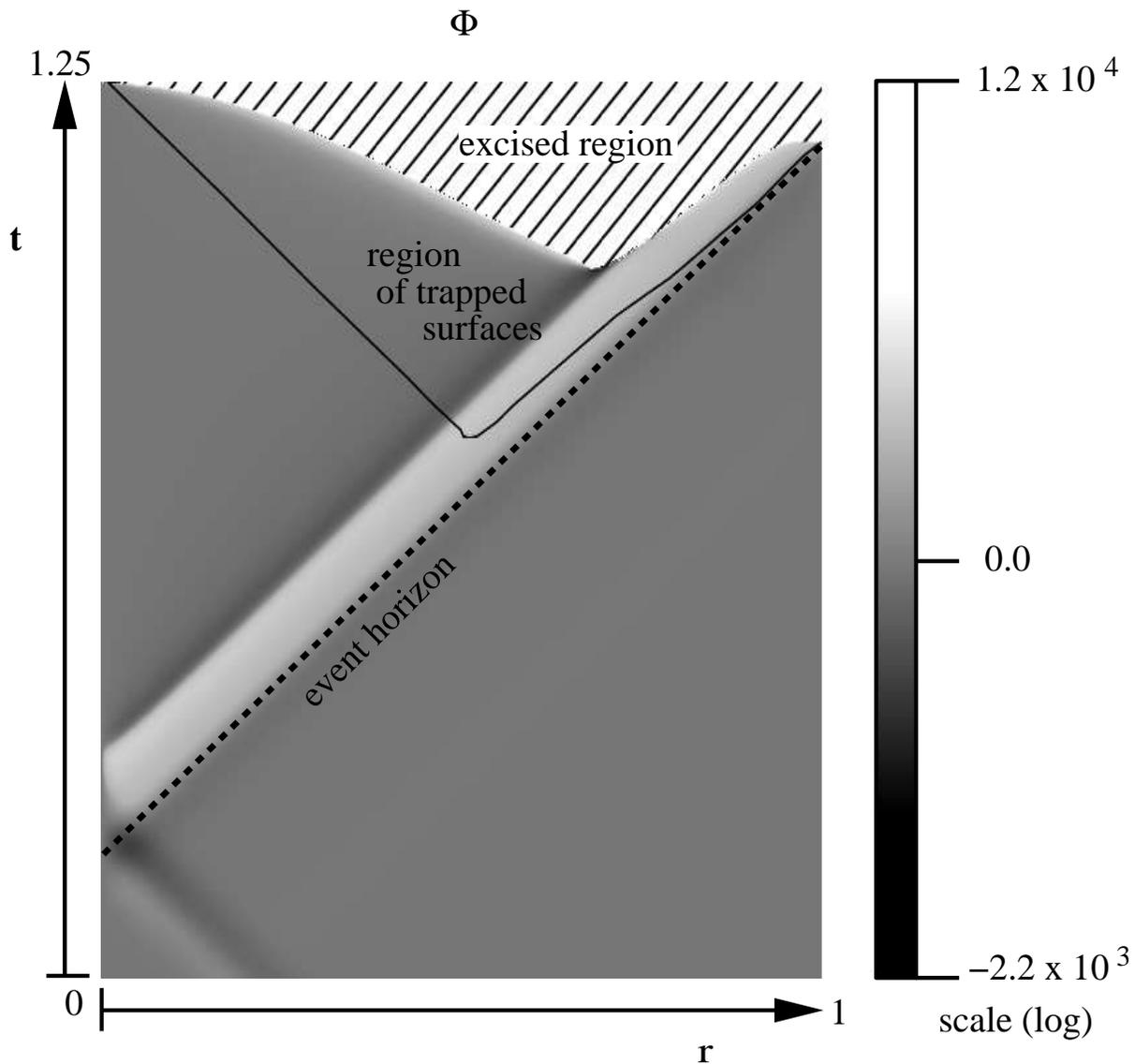}}
\caption{\label{PHI133051_bw}
The gradient of the scalar field $\Phi(r,t)$ on the entire solution domain
for a gaussian with $\amp=0.133051$. On this picture we have also
outlined the region of spacetime containing trapped surfaces, and
drawn in the event horizon with a dashed line (found by tracing a null 
ray backwards in time from the place where the AH reaches $r=1$---which is 
also $\Scri$, so our coordinate system breaks down there). We stop
the simulation at points where the metric variables begin to diverge
(the lower boundary of the excised region),
which presumably is just before a spacetime singularity forms (see Fig
\ref{RS133051_bw} for a similar plot of the curvature scalar). 
}
\end{figure}       

\begin{figure}
\epsfxsize=16cm
\centerline{\epsffile{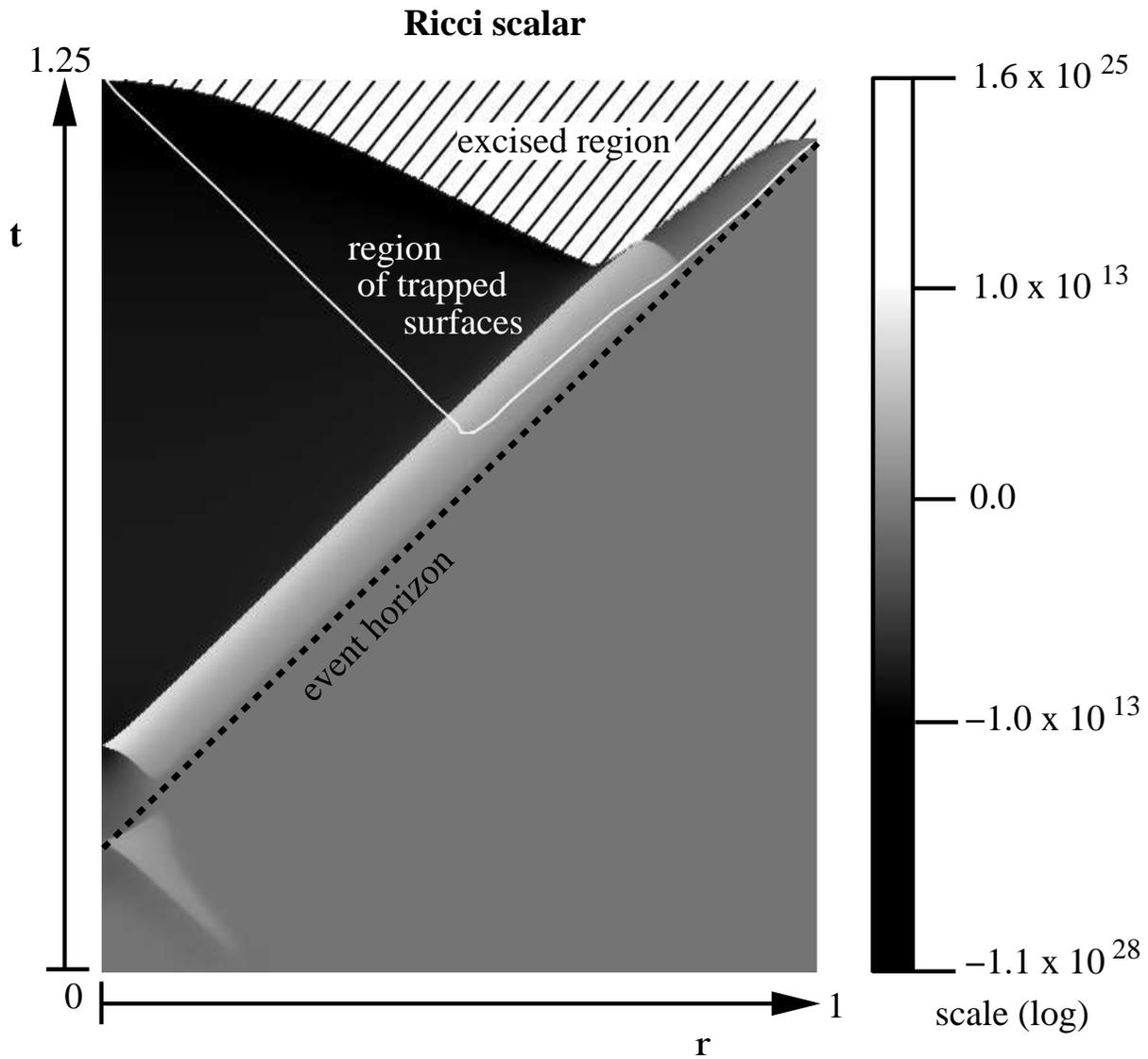}}
\caption{\label{RS133051_bw}
A plot of the Ricci scalar $R(r,t)$ for the same solution as shown in
Fig. \ref{PHI133051_bw}. During most of the evolution $|R|$ is bounded
above by $\approx 10^{13}$, but shortly before reaching the excision
boundary $R$ starts to diverge rapidly, signaling the formation
of a spacelike singularity.
}
\end{figure}

\begin{figure}
\epsfxsize=14cm
\centerline{\epsffile{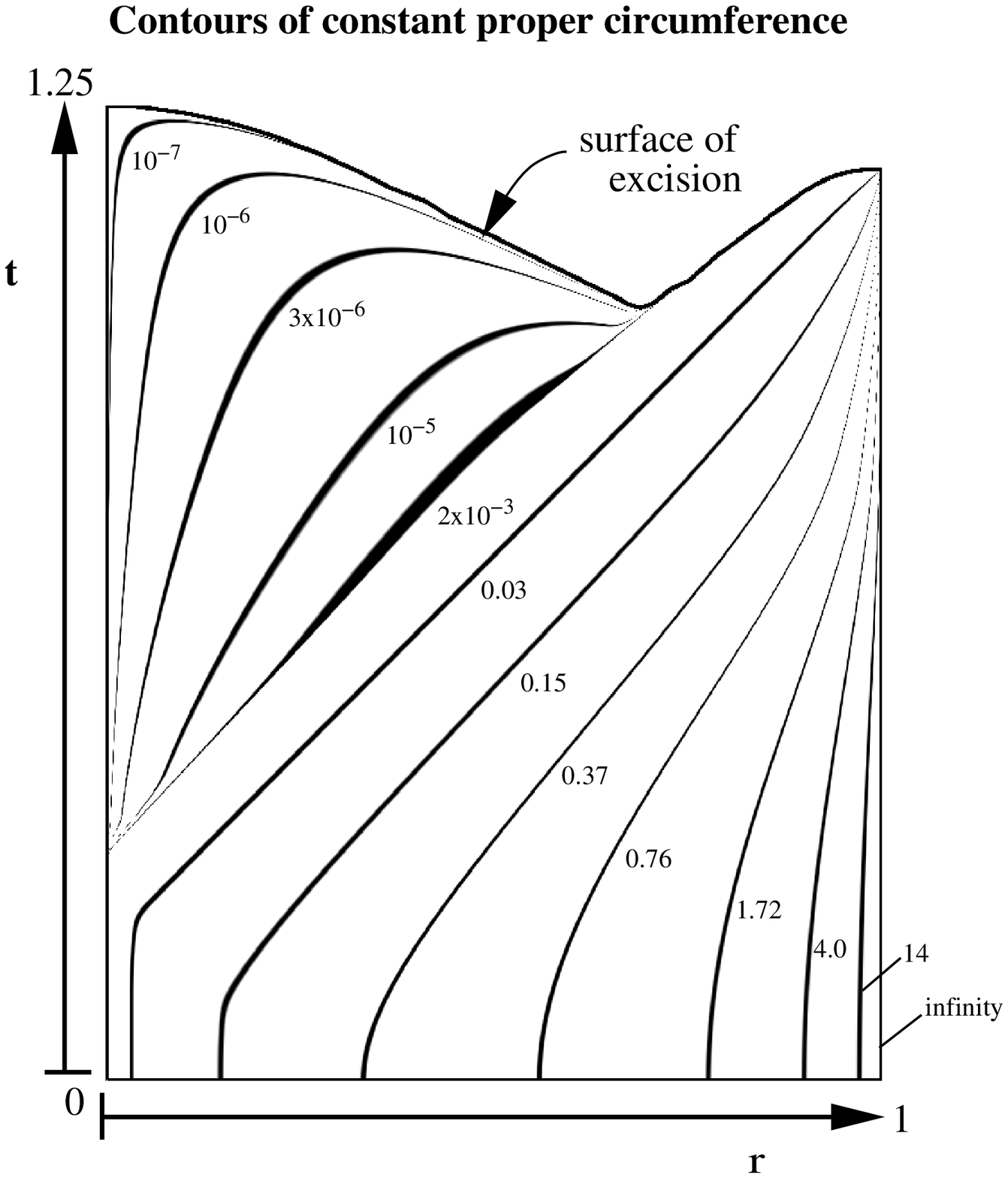}}
\caption{\label{bh13305_prop_bw}
A contour plot of proper circumference (divided by $2\pi$) 
$\rb=\ell\tan(r/\ell)e^B$ for the same solution
as shown in Fig. \ref{PHI133051_bw} (the thickness of each contour line is
constant in units of proper circumference). This plot demonstrates the 
central nature of the singularity. Along the excised surface approaching
$\Scri$ $\rb$ also tends towards zero, though it is not clear with the limited
resolution of this figure there. The event horizon asymptotes to $\rb =
0.037$, i.e. just outside the $\rb=0.03$ contour.
}
\end{figure}

\section{Concluding remarks}

We have studied black hole formation from the collapse of a minimally-coupled
massless scalar field in 2+1 dimensional AdS spacetime. Outside of the event
horizon the spacetime settles down to a BTZ form; in the interior a central, 
spacelike curvature singularity develops. At the threshold of black hole
formation we find that the scalar field and spacetime geometry evolve
towards a universal, continuously self-similar form. 
When a point particle
is present at the origin the critical solution is shifted in central 
proper time by an amount related to the mass of the particle.

By examining the 
behavior of the curvature scalar during sub-critical evolution we deduced
that the universal scaling exponent $\gamma$ for this system 
is roughly $1.2 \pm 0.05$. This value is quite different from the 
scaling exponent $1/2$
derived by Peleg and Steif \cite{shell} for the collapse of thin rings
of dust and by Birmingham and Sen \cite{particles} for particle collisions.
However, those works considered different forms of matter, and the phase
transition was between black hole and naked singularity formation. Thus
one would not expect the same exponent. Also, the local spacetime 
geometry about a dust ring or point-particles is necessarily (empty)
AdS, hence such systems cannot exhibit any of the 
features, other than mass scaling, that are characteristic of 
critical gravitational collapse.

Some questions remain unanswered in this work.
First, what is the exact nature of the critical solution? In other words,
what is the character of the symmetry (if any) responsible for the self-similar 
behaviour, as the system does not appear to admit a global homothetic 
Killing vector \footnote{though, as mentioned in the 
footnote of sec. \ref{sec_css}, David Garfinkle 
has found a CSS solution that is apparently relevant to the AdS critical
solution \cite{garfinkle2}}.
Second, will
{\em any} distribution of energy that could conceivably form a black hole
(i.e. with asymptotic mass $M>0$) eventually do so if one waits long 
enough (because of the Dirichlet boundary conditions imposed on the 
scalar field at $\Scri$)?
A third question, related to the first two, is whether the critical
solution we have found is a true black-hole-formation {\em threshold} solution. 
In other words, that we have a found a universal, CSS solution via
a fine-tuning process indicates that this critical solution is one-mode
unstable; so, does perturbing the critical solution ``one way'' result in a
black hole, and perturbing it the ``other way'' cause the 
scalar field to remain regular, never forming a black hole?
The asymptotic nature of AdS spacetime, 
which is ultimately responsible for the boundary conditions of the scalar field 
at $\Scri$, prevent us from answering this question in our collapse
simulations.

With regards to future work, it would be useful to extend these results
to different scalar-field/geometry couplings, include a mass and potential
terms in the Lagrangian, and to add angular momentum to the initial data to 
study the formation of rotating black holes.
It would also be interesting to understand the critical behavior in light
of the AdS/CFT correspondence. Even though our calculation is purely classical,
there should be a regime where the classical evolution is a good approximation
to the full bulk theory, and consequently there should be a dual CFT
description of the critical phenomena.

\medskip
\noindent
{\bf Acknowledgements} 
We would like to thank David Garfinkle, Viqar Husain, Lior Burko, 
I{\~n}aki Olabarrieta, 
Michel Olivier, Bill Unruh, Jason Ventrella, and Don Witt for many stimulating 
discussions. We are grateful to David Garfinkle for suggesting to us the method 
we used to obtain $\gamma$, as well as the use of the ingoing null 
coordinate system to compare near-critical solutions. 
MWC would particularly like to thank Robert Mann for many early discussions
about this problem during the 1999 {\em Classical and Quantum Physics
of Strong Gravitational Fields} program held at the Institute for
Theoretical Physics, UC Santa Barbara.
This work was supported by NSERC and by NSF
PHY97-22068 and PHY94-07194. Most calculations were 
carried out on the vn.physics.ubc.ca
Beofwulf cluster which was funded by the 
Canadian Foundation for Innovation.

\par

\end{document}